# The Evolution of Turbulent Micro-vortices and their Effect on Convection Heat Transfer in Porous Media

Ching-Wei Huang[1], Vishal Srikanth[1], Andrey V. Kuznetsov[1†]

[1]Department of Mechanical and Aerospace Engineering, North Carolina State University, Raleigh, NC 27695, USA

New insight into the contribution of the microscale vortex evolution towards convection heat transfer in porous media is presented in this paper. The objective is to determine how the microscale vortices influence convection heat transfer in turbulent flow inside porous media. The microscale temperature distribution is analyzed using flow visualization in 2D using streamlines and in 3D using Q structures. The pertinent observations are supplemented with the comparison of surface skin friction and heat transfer using: (1) surface skin friction lines and (2) joint PDF of pressure and skin friction coefficients, along with the Nusselt number. The microscale flow phenomena observed are corroborated with the features of the frequency spectra of the drag coefficient and macroscale Nusselt number. The Large Eddy Simulation technique is used to investigate the flow field inside a periodic porous medium. The Reynolds numbers of the flow are 300 and 500. The porous medium consists of solid obstacles in the shape of square and circular cylinders. Two distinct flow regimes are represented by using the porosities of 0.50 and 0.87. The results show that the surface Nusselt number distribution is dependent on whether the micro-vortices are attached to or detached from the surface of the obstacle. The spectra of the macroscale Nusselt number and the pressure drag are similar signifying a correlation between the dynamics of heat transfer and the microscale turbulent structures. Both vortex shedding and secondary flow instabilities are observed that significantly influence the Nusselt number. The fundamental insight gained in this paper can inform the development of more robust macroscale models of convection heat transfer in turbulent flow in porous media.

## 1. Introduction

Comprehensive studies of the fluid mechanics of turbulence in porous media are only now emerging with the advancements in modern CFD techniques. A fundamental understanding of how turbulent flow influences heat transfer in porous media is still lacking. Turbulence in porous media is different from turbulence in clear fluids because of the restrictions on the size and distribution of turbulent eddies imposed by the finite pore size. The porous medium geometry will determine the properties of turbulence inside it. The repeating solid obstacles will enhance turbulent mixing which promote heat and mass transfer inside the porous media. These unique properties of porous media flows introduce features in microscale flow and thermal transport that are not ubiquitous. For instance, the porous medium imposes a restriction on the length and time scales of turbulence. In this work, microscale refers to a length scale whose order of magnitude is equal to or less than that of the size of the solid obstacle. Macroscale is a mathematically determined scale by applying the Volume Average Theory

† Email address for correspondence: avkuznet@ncsu.edu

(VAT) (Slattery 1967) to the microscale governing equations. Macroscale turbulent structures will not be encountered in this study since the porous medium is assumed to be periodic. This has been demonstrated in the work of Jin *et al.* (2015) and Uth *et al.* (2016). The pore scale prevalence of turbulence has been shown in Jin & Kuznetsov (2017). DNS studies by He *et al.* (2019) verified that the turbulence integral length scale is ~10% of the obstacle diameter in a closely packed porous medium. These observations suggest that the largest turbulent eddies will be formed as the microscale vortices (micro-vortices) behind the solid obstacle.

As a result of the size limitation, turbulence flow phenomena are also limited to the microscale. For example, Chu *et al.* (2018) showed that turbulence kinetic energy is produced near the surface of the solid obstacle at the microscale. It follows from these observations that the size and shape of the solid obstacles will determine the turbulence energy cascade. Convection heat transfer from the solid obstacle surface will also be affected by the flow surrounding the solid obstacle, which is determined by the solid obstacle shape. It is vital to understand the influence of microscale flow structures on turbulent heat transfer in porous media to develop robust macroscale models. The macroscale turbulence models can be used in emerging technologies such as heat management in electronics (Hetsroni *et al.* 2006), long term energy storage systems (Nazir *et al.* 2019), and forest fire modeling (Mell *et al.* 2009). Macroscale turbulence models for porous media flows have followed the Reynolds Averaged, Volume Average (RA-VAT) approach (de Lemos 2012; Lage *et al.* 2007; Vafai 2015; Vafai *et al.* 2009), due to the limited computational power. The macroscale energy models make use of the gradient diffusion hypothesis in conjunction with the assumption of thermal equilibrium between the solid and fluid phases (Nakayama *et al.* 2006). More sophisticated macroscale turbulence models have started to follow the LES approach by including the temporal dynamics of the flow (Wood *et al.* 2020). To the best of the authors' knowledge, such a model for macroscale thermal energy does not exist.

A budget of the macroscale thermal energy equation suggests that macroscale thermal transport is determined by only a few processes. Jouybari *et al.* (2020) noted that macroscale heat transfer is dominated by turbulent convection. Interfacial heat transfer at the microscale plays a critical role in the macroscale thermal energy budget. In the past, interfacial heat transfer has been modeled empirically using Reynolds Averaged Navier-Stokes (RANS) simulations of microscale porous media flow (Kundu *et al.* 2014; Kuwahara & Nakayama 1998; Pedras & de Lemos 2003). Note that the results from microscale RANS simulations are constrained by the modelling error (Iacovides *et al.* 2014) and that these errors are carried forward into the macroscale models based on these RANS simulations. Physical models of the thermal transport in porous media require an understanding of the underlying microscale flow physics.

Previous studies were mostly limited to investigations of microscale turbulent heat convection in tube banks. Wang *et al.* (2006) reported high turbulence intensity in the regions behind the solid obstacles in a staggered tube bank. High turbulence is indicative of enhanced fluid mixing, which could promote heat transfer. Their RANS study has shown that vortex shedding is characterized by a single frequency. This observation can greatly simplify the dynamics of turbulence and heat transfer in porous media. The vortex region behind the heated solid obstacles is associated with a high temperature resulting in low local Nusselt number (Wilson & Bassiouny 2000). Recirculation in the micro-vortices smooths the temperature gradient at the solid obstacle surface for the Reynolds averaged flow (Saito & de Lemos 2006). The link between the micro-vortices and surface Nusselt number has not been investigated. Wilson &

Bassiouny (2000) showed that the Nusselt number increases with in-line tube spacing until a spacing-to-diameter ratio of 3, although the underlying reason was not understood. The LES study by Blackall *et al.* (2020) and the DNS study by Chu *et al.* (2019) provided confirmation of the inhomogeneous distribution of the Nusselt number on the solid obstacle surface, which were previously observed in the RANS studies (Sharatchandra & Rhode 1997). The Nusselt number at the flow stagnation region is more than twice that of the vortex region. LES studies have also shown that the microscale turbulent structures introduce wrinkles in the iso-surface of temperature which in turn influence heat transfer (Linsong *et al.* 2021).

The dynamics of turbulent flow have been studied for two-tandem cylinders with a focus on the vortices that are formed behind the cylinders. The vortex shedding process behind the downstream cylinder of the two-tandem cylinders was reported to be characterized by two frequencies (Alam & Zhou 2008; Zhou & Yiu 2006). The low frequency vortex shedding was caused by the interaction of the vortex shedding behind the upstream cylinder with that of the downstream cylinder. For tandem cylinders, low Reynolds numbers were more conducive for the appearance of two peak frequencies for vortex shedding (Zhou *et al.* 2009). The vortex co-shedding mechanism at high Reynolds numbers was responsible for the disappearance of the second frequency. In addition to the Reynolds number, the separation distance between the cylinders (corresponds to porosity in the present study) also influenced the free shear layers formed behind the cylinder and its interaction with the neighboring cylinder. This interaction is reported to control the transport efficiency of heat and momentum depending on the separation distance between the cylinders. The vortex strength is reported to be a good measure of the heat transfer efficiency, which is controlled by the separation distance. For a low separation distance like typically seen in porous media, the vortices transport heat and momentum with similar efficiency, which is different from an isolated cylinder where heat transport is more efficient. Dual frequency dynamics is also observed in porous media in the present work as shown in section 3.2. However, there is a lack of a systematic study of turbulent flow dynamics and its influence on heat transfer in porous media. It is an important study especially since there are numerous applications of porous media with Reynolds number large enough for the flow to transition to the turbulent regime.

It was shown in the literature that vortices induce intense turbulence mixing, however low Nusselt numbers have also been reported in the vortex region. There is an inadequate understanding of this counterintuitive result. There is also a lack of studies that contrast the heat transfer in different regimes of turbulent flow in porous media. For example, heat transfer properties involving recirculating and shedding vortices are different. The contribution of the different flow regions on the solid obstacle surface to heat transfer has not been studied. Although researchers have reported high Nusselt number in the stagnation region and low Nusselt number in the vortex region, the surface areas covered by these regions are different. The geometry of the porous medium introduces a unique distribution of flow regions. Therefore, the previous observations of the relative contributions of these flow features on heat transfer are inconclusive.

In this paper, we address the shortcomings of the previous research described above. The goal of this study is to determine the influence of the micro-vortices on convection heat transfer in turbulent flow inside porous media. Our preliminary work shows that the turbulent structures inside porous media can either insulate the solid obstacles or carry heat away from them depending on the porous medium geometry (Huang *et al.* 2021, 2019). The micro-vortices will play a vital role in convection heat transfer especially since they are formed on the solid obstacle surface. The porosity of the porous medium and the shape of the solid obstacles both influence the size, shape, and dynamics of the microscale turbulent structures. At high porosity, the size of the micro-vortex scales with the diameter of the solid obstacle. The space in between the solid obstacles is large enough that the micro-vortices are not restricted. At low porosity, the size of the micro-vortex scales with the space in between the solid obstacle surfaces. The formation of a recirculating vortex restricted in between the solid obstacles of porous medium in low porosity is also reported in Linsong *et al.* (2018). The solid surfaces restrict the vortices from growing larger than the pore space (Huang *et al.* 2018). Desai & Vafai (1994) showed that higher gap-width in an annulus geometry decreases the heat transfer rate in natural convection. This suggests that porosity is a critical parameter for heat transfer in the present study. The solid obstacle shape has a higher influence on the turbulent structures when the porosity is low, compared to when the porosity is high. For square cylindrical solid obstacles, slow recirculating vortices were observed that restrict heat transfer. For circular cylindrical solid obstacles, the vortices were able to shed into the primary flow, dissipating heat during the process (Huang *et al.* 2019). We note that an effective way of increasing the heat transfer rate from bluff objects is to increase the Reynolds number of the flow as shown by Khanafer & Vafai (2021). In order to change heat transfer characteristics for a given Reynolds number, either the properties of the solid obstacle shape or the fluid can be changed.

Establishing a direct connection between the micro-vortices and the heat transfer is a novel contribution of the present work. Understanding the properties of the microscale flow is important for the macroscale heat transfer modeling as well. Microscale studies have shown that the heat transfer efficiency between the solid obstacle surface and the fluid increases with an increase in the Reynolds number and obstacle diameter (Chu *et al.* 2019; Suga *et al.* 2017). The microscale simulations for square rods (Kuwahara & Nakayama 1998), circular rods (Rocamore Jr. 2001), and elliptic rods (Pedras & de Lemos 2008) revealed that the thermal dispersion varies drastically with the solid obstacle shape. The functional dependence of the Nusselt number on porosity changes with the solid obstacle shape (Torabi *et al.* 2019). High Resolution Large Eddy Simulation (LES) studies of finite pebble beds show that hot spots that appear on the surface of the pebbles are highly unsteady, in which their locations move over time (Shams *et al.* 2014). Turbulent thermal mixing for circular rods increases with an increase in the Reynolds number and approaches an asymptotic value at higher Reynolds numbers (Li & Wu 2013). Therefore, heat transfer in porous media is highly unsteady and closely linked to the formation, propagation, and dissipation of flow structures. For instance, the temperature fluctuation intensity observed in the present work is ~15% in the vortex region for a porous medium with a porosity of 0.50. The fluctuations come from vortex shedding and flow instabilities (discussed in section 3). These time dependent effects are important when simulating turbulent convection heat transfer in porous media.

In this paper, we focus on investigating the relation between microscale turbulent structures dynamics and heat transfer in forced convection in porous media. We used LES to simulate the microscale flow inside of a periodic porous medium consisting of an array of cylinders. The simulation conditions and the details of the numerical method are discussed in section 2. We used a multiscale approach to analyze the flow in section 3. We used the flow streamlines and coherent structure visualization using the Q- criterion to visualize the vortex dynamics and overlaid the temperature distribution to determine its influence on heat transfer. We analyzed the dynamics of the micro-vortices and heat transfer at the macroscale using a frequency transformation of the time series of the drag coefficient ($C_D$), lift coefficient ($C_L$) and Nusselt number ($Nu_m$). We identified the relation between the surface heat transfer and shear using the surface skin friction lines, Nusselt number distributions, and joint Probability Density Functions (PDFs) on the solid obstacle surface. Using these techniques, we showed that the microscale turbulent structures dynamics directly influences heat transfer in forced convection in porous media.

## 2. Solution Methods

### 2.1. Simulation Geometry and Boundary Conditions

A homogeneous porous medium is constructed by arranging cylindrical solid obstacles in a simple square lattice (see figure 1). The simulation domain is three-dimensional spanning 4 unit cells along the *x*- and *y*- directions, and 2 unit cells in the *z*- direction. The dimensions are chosen based on the decorrelation width for turbulence two-point velocity correlation functions from Jin *et al.* (2015) and Uth *et al.* (2016). This forms a Representative Elementary Volume (REV) of size $4s \times 4s \times 2s$, where $s$ is the pore size (figure 1). In Appendix A, it is shown that the REV of size $4s \times 4s \times 2s$ is sufficient to calculate the macroscale quantities. The simulations were performed using the commercial CFD code ANSYS Fluent 16.0. The details of the numerical method presented in section 2 are taken from the Fluent Theory Guide (ANSYS Inc. 2016). Periodic boundary conditions are used to impose an infinite span in all directions to represent a homogenous porous medium. The pressure and velocity variables are periodic in all three Cartesian directions. The no-slip boundary condition is imposed at the solid obstacle surfaces. The temperature variable is periodic in the *y*- and *z*- directions. A temperature gradient is imposed in the *x*- direction such that the bulk temperature at the inlet of the REV is 323 K by following the methodology given in the Fluent Theory Guide (ANSYS Inc. 2016). The temperature of the solid obstacle surfaces ($T_w$) was set to a constant value of 353 K. This results in a characteristic temperature difference $\Delta T$ of 30 K. For this temperature change, we do not expect any changes in the physical state of the fluid or the occurrence of chemical reactions. The independence of the Nusselt number distribution to the characteristic temperature difference near the chosen value of 30 K is shown in Appendix B. The Prandtl number ($Pr$) was kept constant at 7. The fluid used for simulation is water (properties listed in table 1). Two values of porosity ($\varphi$), 0.50 and 0.87, are studied to represent two different regimes of turbulent flow in porous media. The flow features at $\varphi = 0.50$ are similar to that of internal flow (such as channel flow), while the flow at $\varphi = 0.87$ resembles external flow (such as flow around a bluff body). At $\varphi = 0.50$, the solid obstacle surfaces are close to each other such that the separated shear layer bridges the gap between the two obstacles. A channel-like flow is observed in between the recirculating vortices. The recirculating vortices are also driven by the shear layer in a manner that is similar to lid-driven cavity flows. At $\varphi = 0.87$, the solid obstacle surfaces are far apart such that the vortices are able to form, detach and get transported away from the solid obstacle surface. The Reynolds number ($Re_p$) is 300 for a majority of the cases. A single case with a Reynolds number of 500 is used to understand how Reynolds number affects the observations put forth in this paper. The square solid obstacle shape is used at high Reynolds number to avoid the deviatory flow observed in Srikanth *et al.* (2021). The flow rate was sustained by using a constant applied pressure gradient ($\rho g_1$) as the driving force. The porosity ($\varphi$) and the Reynolds number are defined as,

$$\varphi = 1 - \frac{\pi}{4}\left(\frac{d}{s}\right)^2 \tag{2.1}$$

$$Re_p = \frac{u_m d}{\nu} \tag{2.2}$$

where $d$ is the hydraulic diameter of the solid obstacles, $u_m$ is the superficially averaged macroscale mean velocity in the *x*- direction, and $\nu$ is the kinematic viscosity of the fluid. LES with the Dynamic One-equation Turbulence Kinetic Energy (DOTKE) subgrid model (Kim & Menon 1997) is used to simulate the microscale flow field inside the porous medium. Rodi (1997) demonstrated the superior performance of LES in simulating bluff body flows compared to RANS. Dynamic LES models are able to reproduce experimental results with reasonable accuracy at a fraction of the cost of DNS (Jin *et al.* 2016). Davidson & Krajnovic (2000) have also demonstrated the ability of the DOTKE model to predict parameters associated with vortex shedding. The simulations have been run on the North Carolina State University Linux Cluster. Representative computation time for one LES case is 25,000 CPU-Hours (1 CPU-Hour = Computation Time in Hours for a single CPU). Each simulation is run using 80 cores with nodes consisting of Intel® Xeon® E5-2650 v4 processors. Each simulation is run for 200 non-dimensional time units (8,000 CPU-Hours) to equilibrate the solution followed by the collection of turbulence statistics for 400 non-dimensional time units (17,000 CPU-Hours).

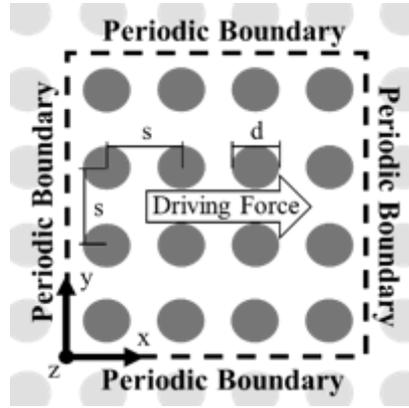

Figure 1: The Representative Elementary Volume (REV) geometry of a porous medium. The distance between the centers of the solid obstacles *s* (pore size) and the solid obstacle diameter *d* are also shown in the figure.

| | |
|---|---|
| Density $\rho$ $(kg/m^3)$ | 998.2 |
| Specific Heat $C_p$ $(J/kg \cdot K)$ | 4182 |
| Thermal Conductivity $k$ $(W/m \cdot K)$ | 0.6 |
| Dynamic Viscosity $\mu$ $(kg/m \cdot s)$ | 0.001003 |

Table 1: The properties of the fluid used in simulation corresponding to water at 323 K.

### 2.2. Details of the Physical Model and Numerical Method

The governing equations for the LES are the filtered Navier-Stokes equations (2.3) – (2.4). The tilde notation ($\tilde{u}$) denotes the spatial filtering operator. The pressure variable $\tilde{p}$ here is the filtered periodic pressure (terminology adopted from the ANSYS Inc. 2016 theory guide). The pressure gradient term in the governing equations in periodic flows can be split into a constant pressure gradient term $\rho g_i$ and the gradient of the periodic pressure $\partial \tilde{p}/\partial x_i$. To calculate the static pressure, we take the sum of the periodic pressure and the linearly varying pressure. The subgrid velocity scale is estimated by solving a transport equation for the subgrid turbulence kinetic energy $k_{SGS}$ (equation (2.5)). The subgrid scale filter length $\Delta$ is set as the cube root of the cell volume. Kim (2004) has demonstrated that setting the grid filter length as the cube root of the cell volume yields accurate results using unstructured, stretched grids for simulating the flow inside channels and around square cylinders. We note that there are other procedures used to calculate the filter width such as the maximum side length of the hexahedral cell or by using the van Driest damping function. Equation (2.6) estimates the subgrid turbulence eddy viscosity. The characteristic subgrid length scale for the calculation of subgrid turbulent viscosity is estimated as $\Delta$. The model constants $C_k$ and $C_\varepsilon$ are determined by the localized dynamic subgrid-scale model from Kim & Menon (1997). The grid scale velocity field is filtered to a test scale velocity field. The test filter length $\hat{\Delta}$ is equal to twice the grid filter length $\Delta$ (ANSYS Inc. 2016). The similarity between the stresses at the two scales is invoked to determine the model constants. The model constant $C_k$ is determined in equations (2.7) – (2.8) by using the similarity between the SGS stress tensor $\tau_{ij}$ and the test Leonard stress tensor $L_{ij}$. The value of $C_k$ is limited by $-\mu/(k_{SGS}^{1/2}\Delta)$ to avoid a negative total viscosity. Model constant $C_\varepsilon$ is determined in equation (2.9) by using the similarity between the dissipation rate at the grid level $\varepsilon_{SGS}$ and the test level $\varepsilon_{test}$. The governing equations of thermal energy are given in equations (2.10) – (2.12). The turbulent Prandtl number ($Pr_T$) is assumed to take a constant value of 0.85. Dynamic methods are available for the calculation of turbulent Prandtl number in a manner using the Germano identity at the cost of added complexity in the numerical model. It is noted in the literature that the use of a variable turbulent Prandtl number has a negligible effect on the prediction of thermal characteristics in wall bounded flows when compared to using the constant value of 0.85 (Kakka & Anupindi 2020). The filtered governing equations are solved in conjunction with the DOTKE subgrid model using the Finite Volume Method (FVM). A box filter is implicitly applied by the computational grid in the FVM.

$$\frac{\partial \widetilde{u_j}}{\partial x_j} = 0 \tag{2.3}$$

$$\frac{\partial \rho \widetilde{u_i}}{\partial t} + \frac{\partial \rho \widetilde{u_i}\widetilde{u_j}}{\partial x_j} = -\frac{\partial \tilde{p}}{\partial x_i} + \frac{\partial}{\partial x_j}\left[(\mu + \mu_{T,SGS})\left(\frac{\partial \widetilde{u_i}}{\partial x_j} + \frac{\partial \widetilde{u_j}}{\partial x_i}\right)\right] + \rho g_i \tag{2.4}$$

$$\frac{\partial k_{SGS}}{\partial t} + \frac{\partial (\widetilde{u_j} k_{SGS})}{\partial x_j} = \left[C_k k_{SGS}^{\frac{1}{2}} \Delta \left(\frac{\partial \widetilde{u_i}}{\partial x_j} + \frac{\partial \widetilde{u_j}}{\partial x_i}\right)\right]\frac{\partial \widetilde{u_i}}{\partial x_j} - C_\varepsilon \frac{k_{SGS}^{\frac{3}{2}}}{\Delta} + \frac{\partial}{\partial x_j}\left(\mu_{T,SGS}\frac{\partial k_{SGS}}{\partial x_j}\right) \tag{2.5}$$

$$\mu_{T,SGS} = C_k k_{SGS}^{\frac{1}{2}} \Delta \tag{2.6}$$

$$\tau_{ij} = -2C_k k_{SGS}^{\frac{1}{2}} \Delta \widetilde{S_{ij}} + \frac{2}{3}\delta_{ij} k_{SGS}; \quad L_{ij} = -2C_k k_{test}^{\frac{1}{2}} \hat{\Delta} \widehat{\widetilde{S_{ij}}} + \frac{1}{3}\delta_{ij} L_{kk} \tag{2.7}$$

$$C_k = \frac{1}{2}\frac{L_{ij}\sigma_{ij}}{\sigma_{ij}\sigma_{ij}}; \quad \sigma_{ij} = -\hat{\Delta} k_{test}^{\frac{1}{2}} \widehat{\widetilde{S_{ij}}}; \quad k_{test} = \frac{1}{2}\left(\widehat{\widetilde{u_k}\widetilde{u_k}} - \widehat{\widetilde{u_k}}\widehat{\widetilde{u_k}}\right) \tag{2.8}$$

$$C_\varepsilon = \frac{\widehat{(\partial \widetilde{u_i}/\partial x_j)(\partial \widetilde{u_i}/\partial x_j)} - (\partial \widehat{\widetilde{u_i}}/\partial x_j)(\partial \widehat{\widetilde{u_i}}/\partial x_j)}{\left((\mu + \mu_{T,SGS})\hat{\Delta}\right)^{-1} k_{test}^{\frac{3}{2}}} \tag{2.9}$$

$$\frac{\partial \rho E}{\partial t} + \frac{\partial (\rho E + \tilde{p})\widetilde{u_j}}{\partial x_j} = \frac{\partial}{\partial x_j}\left[(k_{eff})\frac{\partial \tilde{T}}{\partial x_j}\right] + \frac{\partial}{\partial x_j}\left[\widetilde{u_j}(\mu + \mu_{T,SGS})\left(\frac{\partial \widetilde{u_i}}{\partial x_j} + \frac{\partial \widetilde{u_j}}{\partial x_i}\right)\right] \quad (2.10)$$

$$E = C_p(\tilde{T} - 298.15) + \frac{1}{2}\widetilde{u_j}^2 \quad (2.11)$$

$$k_{eff} = k + k_T; \text{ where } k_T = \frac{\mu_{T,SGS} C_p}{Pr_T} \text{ and } Pr_T = 0.85 \quad (2.12)$$

The bounded second-order central scheme (according to the work of Leonard 1991) is used to discretize the convective terms and a second-order central scheme for the viscous terms in the momentum equation. The thermal energy equation is discretized using the QUICK scheme for increased stability without compromising accuracy. The locations of the pressure and velocity variables are staggered. The pressure is stored at the centroid of the face of the cell, while the velocity is stored at the cell center. The momentum and pressure Poisson equations are solved in a segregated manner using a pressure-implicit scheme with splitting of operators (PISO). The thermal energy equation is solved at the end of the time step. The simulation is advanced in time using a second-order implicit backward Euler method. The no-slip boundary condition is enforced at the solid obstacle walls by specifying the surface tangential and normal velocities at the solid obstacle wall to be equal to zero. We note that there are other methods of implementing the no-slip boundary condition at the wall, such as mirroring or by polynomial reconstruction of the fluid velocity distribution in the solid obstacle domain for accurate gradient representation. In this work, we are modeling only the fluid domain and there are no nodes inside the solid obstacle to model the solid phase. Therefore, we explicitly specify velocity boundary conditions at the solid surface. For the periodic boundary conditions, the grid point locations are conformal for each pair of periodic faces. Each pair of periodic faces is treated as a connected interface with connected nodes to periodically repeat the velocity and pressure distributions. A similar procedure is followed for the temperature distribution in the *y*- and *z*- periodic faces. For the *x*- periodic faces, the temperature distribution at the exit face is copied to the inlet face after adjusting the magnitude to specify the desired bulk inlet temperature.

*2.3. Validation of the Physical Model and Numerical Method*

In this section, we demonstrate that the numerical method described in section 2.2 is appropriate for simulating turbulent flow in porous media by comparing the simulated flow field with that of an experimental study. The ANSYS Fluent code used in this paper is well-validated for canonical turbulent flows (Fluent Inc. 2006). However, the code must be validated for turbulent flows in porous media since the features of both flows inside channels and flows around a single solid obstacle are observed. The experimental results of Aiba *et al.* (1982) for turbulent flow through a tube bank in a channel are viable for validation due to the small size of the tube bank. A limitation of this comparison is that the Reynolds number of the flow used in the experimental work is an order of magnitude larger than that used in the present work. The computational grid used in the validation simulation is capable of capturing the large-scale eddies. The DOTKE sub grid scale model is able to predict the small-scale eddies. At high Reynolds numbers, a larger portion of the turbulence energy spectrum is predicted by the subgrid model than that at low Reynolds numbers. Therefore, the use of a large Reynolds number for validation is beneficial to determine the performance of the LES subgrid model. An excellent agreement between the simulation and experiment at the high Reynolds number implies an even better simulation accuracy at low Reynolds numbers. This is because the subgrid flow satisfies the subgrid model assumptions at low Reynolds numbers.

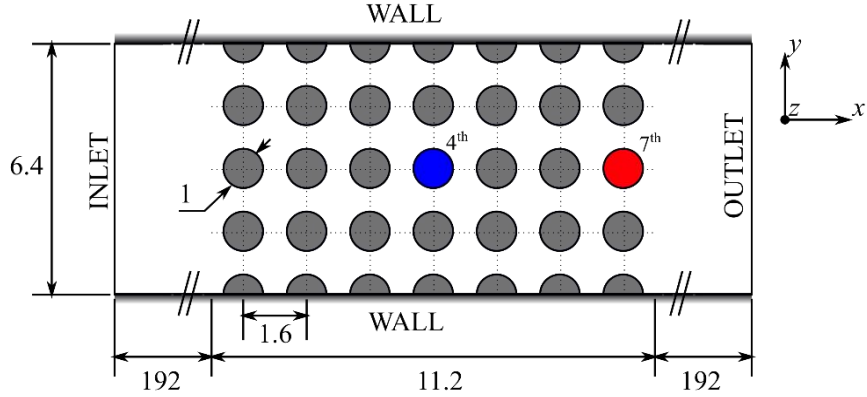

Figure 2: A sketch of the computational domain used to reproduce the experimental results of Aiba *et al.* (1982) for validation of the numerical method. The simulations are performed to compare the coefficient of pressure and Nusselt number on the colored tubes shown in the figure with that of the experiment. All the lengths are non-dimensionalized using the tube diameter.

The geometry used for the validation simulation is shown in figure 2. The validation simulation models the experimental case with a dimensionless tube spacing of 1.6 and a Reynolds number of 41,000. The Reynolds number is calculated using the mean flow velocity at the minimum clearance and the tube diameter as per Aiba *et al.* (1982). Note that all the lengths are non-dimensionalized using the tube diameter. The fluid properties are that of air at 25°C. The following approximations are made while modeling the experimental setup:

1. The flow at the center of the tube span is modeled by introducing a periodic boundary condition in the *z*- direction. The approximation follows from the nearly constant velocity distribution in the middle of the channel for turbulent flow. The span of the periodic domain in the *z*- direction is two times the pore size. The turbulence two-point correlation width is less than the span of the domain.

2. Sufficiently long entrance and exit sections to the test section are introduced such that the flow becomes fully developed. The entrance and exit sections of the computational domain are 30 times the channel width.

3. Grid stretching is used to increase the grid size at the inlet and outlet since high grid resolution in these regions is not important to the flow in the test section.

4. The constant velocity boundary condition is specified at the inlet. Spectrally synthesized perturbations are imposed on the velocity inlet to simulate a 5% turbulence intensity as per the work of Aiba *et al.* (1982). Atmospheric pressure is specified at the outlet.

5. In the experiment, only the tube that is being considered for measurement is heated at a given time. We follow the same procedure in the simulation by assigning a constant heat flux boundary on the surface of a single cylinder. The experimental heat flux has not been reported in the original article. We have assumed the heat flux to be equal to 2 W/m$^2$ following the experimental work of Sarma & Sukhatme (1977) for the flow

   over a single cylinder. The induced temperature increase is low enough that it does not violate the assumptions of the physical model used for the simulation.

6. In the experiment, the heat transfer measurements correspond to a more complex conjugate heat transfer problem that considers both the effects of the solid and the fluid phases. In the simulation, we are assuming that the solid surfaces that are not heated are adiabatic.

The LES model equations used for the simulation are described in section 2.2. The distributions of the coefficient of pressure ($C_{pressure}$) on the surface of the 4$^{th}$ and 7$^{th}$ tubes are used for comparison (figure 3(a)). The $C_{pressure}$ is calculated as per the definition given by Aiba *et al.* (1982). The simulated $C_{pressure}$ distribution follows the same trend as that of the experiment. The simulated stagnation pressure is less than that of the experiment. In the low pressure regions, the quantitative agreement between the simulation and the experiment is excellent. The disparity between the simulation and the experiment is due to turbulence model limitations, coarse grid resolution, and differences between the simulation and experimental setup.

The distribution of the Nusselt number on the surface of the tubes is shown in figure 3(b). There exists an average error of 15% between the simulation and the experimental Nusselt number distributions. It should be noted that Iacovides *et al.* (2014) reported close to zero average error in the Nusselt number distribution with the experimental result of Aiba *et al.* (1982) by assuming periodicity in the flow in all 3 directions and that all of the cylinders are heated. In the present work, we observe an excellent qualitative agreement in the distribution of the Nusselt number. The distribution of the Nusselt number is virtually an exact match between the simulation and the experiment after adjusting for the mismatch in the magnitude of the Nusselt number. This suggests that the simulation adequately captures the features of the flow at the solid obstacle surface and the dependence of heat transfer on these flow features. There are several differences between the simulation and the experiment that could have led to the quantitative difference in Nusselt number:

1. The incomplete specification of the wind tunnel inlet and outlet conditions in the simulation geometry.
2. The assumption of an infinitely periodic span for the tubes.
3. The assumption of adiabatic tunnel and tube walls and the lack of a conjugate heat transfer model in the simulation to consider the heat transfer inside the solid tube.

Sources of experimental error could also be a contributing factor. The margin of error in the temperature measurements is not reported in the original paper, but similar experiments report a 3% margin of error in the temperature measurements. Errors in the temperature value from both the experiment and the simulation get accentuated during the calculation of the Nusselt number since the inverse of temperature difference is considered.

Excellent quantitative agreement of $C_{pressure}$ in the low pressure region demonstrates the ability of the model to predict the vortex region, which is the primary focus of the paper. We also note that the coarse grid resolution is adequate to capture the main features of the flow. Resolution up to the Kolmogorov scale will not contribute new information in this study. Therefore, we have established that the numerical method described in section 2.2 is able to reproduce the flow behavior in porous media that is observed in experimental work. We are proceeding to use the numerical method for our analysis. The numerical accuracy will improve further when compared to the validation case due to the high-resolution grids and the low Reynolds number used in the present work.

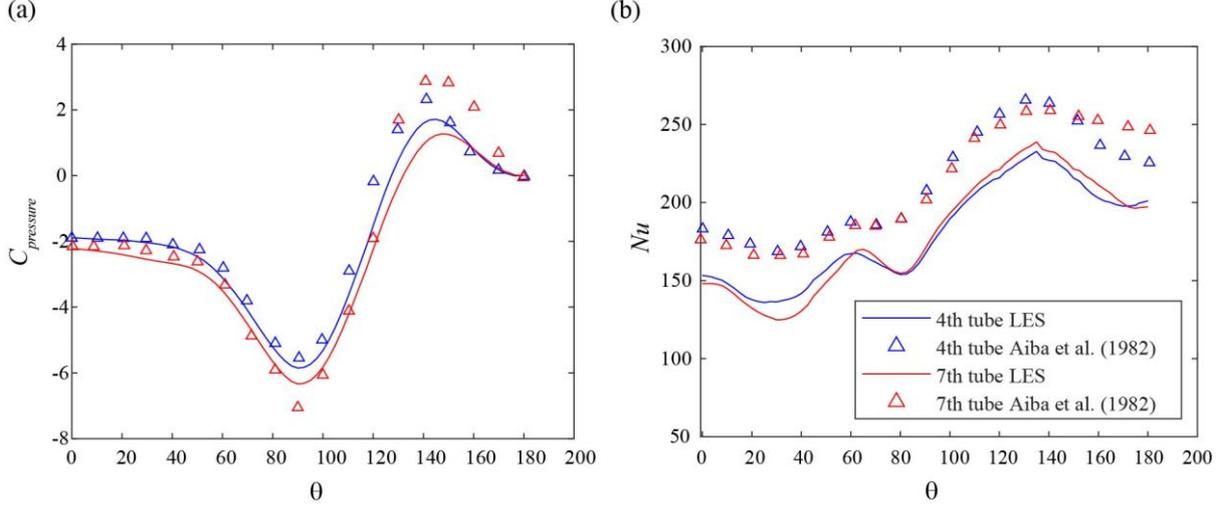

Figure 3: The distribution of the (a) coefficient of pressure, and (b) Nusselt number on the surfaces of tubes 4 and 7 in the center row of the tube bank (figure 2) for the LES and the experiment.

### *2.4. Choice of Grid Resolution*

To determine a suitable grid resolution for the simulations, we perform a grid convergence study using an REV size of $2s \times 2s \times 2s$ since the study concerns the smallest scales. The grids used in this work are unstructured and follow a block structured O-grid topology that stretches the grid around the solid obstacle surface. The grid cells in the bulk of the computational domain are cubical in shape with an aspect ratio of ~1. The grid cells at the solid obstacle surface have a grid step size that is one order of magnitude smaller than the maximum grid step size. The clustering of the cells at the solid obstacle surface are to accurately capture the boundary layer. The maximum value of the grid spacing $\Delta x_{max}$ and the non-dimensional near-wall grid spacing $\Delta y^+_{max}$ are shown in table 2. The distribution of $\Delta y^+$ on the solid obstacle surface is shown in figure S1 in the supplementary material. First, the LES Index of Quality (LES_IQ) (Celik *et al.* 2005) is used to estimate the fraction of the total turbulence kinetic energy that is resolved by the grid. Pope (2004) recommends resolving 80% of the turbulence kinetic energy for LES. Remarks from Celik *et al.* (2005) indicate that simulations may be considered to be of DNS quality with LES_IQ > 0.9. The volume weighted average of the LES_IQ for all the LES simulations in this work is greater than 0.8. The minimum and volume averaged values of LES_IQ at an instant in time are reported in table 3. LES_IQ values less than 0.8 are observed in the near-wall region for the instantaneous flow as spots on the solid obstacle surface. The location of these spots coincides with the impingement of turbulent structures on the solid obstacle surface. This implies an increased reliance on the subgrid model in the near-wall region. The subgrid model has been validated in section 2.2 and its performance has been deemed adequate.

Next, the turbulence kinetic energy spectrum is calculated using the one-dimensional turbulence two-point velocity fluctuation correlation functions. The turbulence kinetic energy spectrum is used to confirm that the large-scale eddies and a substantial portion of the inertial subrange have been resolved in this work. The turbulence kinetic energy spectra ($E_{ii}/3$) versus the non-dimensional wavenumber ($ks$) for the LES test cases are shown in figure 4. A portion of the turbulence energy spectrum aligns with -5/3 slope for all the cases, which indicates that the inertial subrange has been captured. At the small scales of turbulence, the turbulence kinetic energy declines by three orders of magnitude when compared to the largest scales. Therefore, the smallest scales of turbulence are not significant in our study. An experimental study by Nguyen *et al.* (2019) shows a lack of small scale coherent structures in randomly packed porous media. This further supports the notion that the small length scale motions do not contribute significantly to surface heat transfer. In some cases, a portion of the energy spectrum at the high non-dimensional wave numbers is excited due to numerical noise. This is also observed in the energy spectrum plots shown in Eggels *et al.* (1994). Comparing the turbulence energy spectrum produced by the three grid sizes, the grid sizes $\Delta x_{max}/s$ of 0.025 and 0.0125 show similar trends for a wider range of the length scales until the dissipative scales of turbulence are reached. Note that the spectra will not be coincident due to the limitations of Discrete Fourier Transforms that introduces oscillations in the spectra that are unique to each case. Since we are interested in the trends in the spectra rather than an exact point-by-point comparison, the oscillations do not impact our study. Since we are using a subgrid model for the dissipative scales, we are proceeding with the analysis using a grid size $\Delta x_{max}/s$ of 0.025.

| Porosity $\varphi$ | Solid obstacle shape | Near-wall grid size | Coarse grid, $\Delta x_{max}/s=$ 0.05 | Intermediate grid, $\Delta x_{max}/s=$ 0.025 | Fine grid, $\Delta x_{max}/s=$ 0.0125 |
|---|---|---|---|---|---|
| 0.50 | Circle | 0.0022$s$ | 1.06 | 1.16 | 1.05 |
| 0.87 | Circle | 0.0075$s$ | 1.68 | 1.81 | 1.79 |
| 0.87 | Square | 0.0044$s$ | 1.87 | 1.65 | 1.79 |

Table 2: The maximum value of non-dimensional near-wall grid spacing, $\Delta y^+_{max}$, calculated on the surface of the solid obstacles for the grid resolution test cases. These are small areas with high $\Delta y^+$ values, overall $\Delta y^+$ values on the solid obstacle surfaces are kept below 1.

| Porosity $\varphi$ | Solid obstacle shape | | Coarse grid, $\Delta x_{max}/s=$ 0.05 | Intermediate grid, $\Delta x_{max}/s=$ 0.025 | Fine grid, $\Delta x_{max}/s=$ 0.0125 |
|---|---|---|---|---|---|
| 0.50 | Circle | Minimum | 0.22 | 0.44 | 0.66 |
|      |        | Average | 0.81 | 0.95 | 0.98 |
| 0.87 | Circle | Minimum | 0.70 | 0.73 | 0.80 |
|      |        | Average | 0.96 | 0.98 | 0.99 |
| 0.87 | Square | Minimum | 0.69 | 0.74 | 0.89 |
|      |        | Average | 0.95 | 0.98 | 0.99 |

Table 3: The value of LES_IQ calculated for the grid resolution test cases. Both the minimum and the volume-averaged values are reported (ranges from 0 to 1, values close to 1 indicate high resolution with a large fraction of the turbulence kinetic energy being resolved).

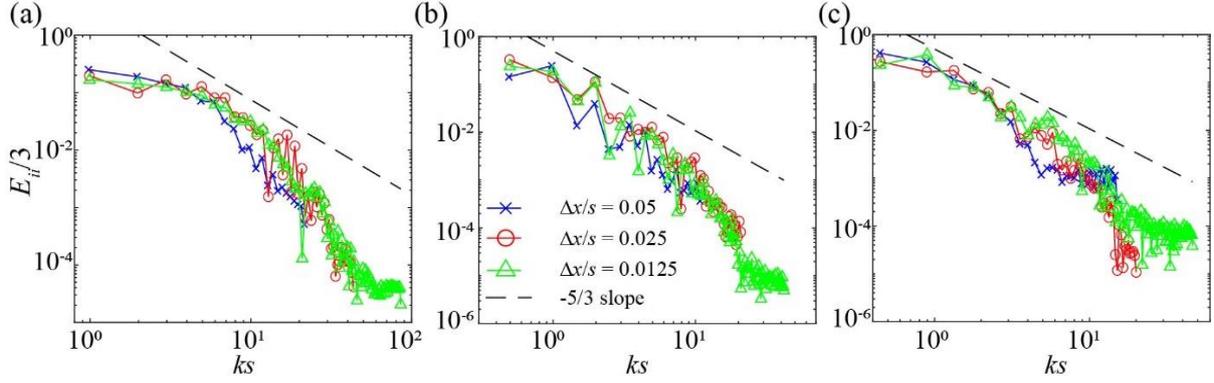

Figure 4: Turbulence kinetic energy $E_{ii}/3$ versus the non-dimensional wavenumber $ks$ for LES cases: (a) $\varphi = 0.50$ with circular solid obstacles; (b) $\varphi = 0.87$ with circular solid obstacles; (c) $\varphi = 0.87$ with square solid obstacles. The dashed line corresponds to the -5/3 slope on the log plot.

## 3. Results and Discussion

The investigation of the turbulent flow physics and heat transfer inside a porous medium requires a multiscale analysis with an emphasis on the microscale flow behavior. In this section, we are analyzing the microscale vortex transport and its effect on heat transfer using the following approach.

1. First, we identify the regions inside the REV that contribute significantly to heat transfer by visualizing the 3D flow field. We visualize the flow patterns using the instantaneous flow streamlines and the 3D coherent structures using the Q-criterion. We identify significant regions of heat transfer by locations with a high temperature gradient. We use skin friction lines overlaid on the Nusselt number distribution on the solid obstacle surfaces to investigate the influence of the surface shear on heat transfer.
2. With an understanding of how the different flow features observed in porous media flow influence the heat transfer, we investigated the dynamics of the surface-averaged flow properties. This helps reduce the complexity of the analysis when compared to a high frame rate visualization of the 3D flow structures. We use the Fast Fourier Transform of the time series of lift coefficient ($C_L$), drag coefficient ($C_D$), and Nusselt number ($Nu_m$) to transform them into the frequency domain and correlate them with our observations from the 3D flow visualization. We determine whether the spectra of $C_L$, $C_D$, and $Nu_m$ have any similarities.
3. Next, we determine the fractional contribution of the different flow features towards heat transfer. We constructed a joint PDF of the surface pressure and skin friction coefficients with the surface Nusselt number to map surface processes on the solid obstacle. We have also used it to test our hypothesis that the pressure and shear dominated processes during stages of vortex shedding have high impact on heat transfer.

We have applied this approach to the simulation cases listed in table 4. We discuss our results in the following sections.

### *3.1. Visualization of the turbulent structures and temperature distribution*

Our previous work suggests that the micro-vortices thermally insulate the solid obstacle since the vortices have a smaller velocity than the surrounding flow (Huang *et al.* 2021, 2019). The heat transfer rate of the obstacle surface increases as the contact area between the micro-vortices and the solid obstacle decreases. This suggests that the micro-vortices reduce the effective surface area that is available for heat transfer. These inferences need to be supplemented by Large Eddy Simulations of the vortex dynamics, since only the Reynolds averaged flow field was considered in the past. In the present work, we have identified that two types of vortex systems can occur in porous media, namely the recirculating (figure 5(a)) and shedding vortices (figure 5(b-d)). The dynamic characteristics of the two vortex systems are distinct, and they impart unique attributes to the heat transfer as shown in section 3.2. The occurrence of each vortex system depends on the porous medium geometry, primarily the porosity. The stark contrast in the dynamics of the two vortex systems warrants a time dependent analysis, which is the methodology adopted in this paper.

| Case ID | Reynolds number | Solid obstacle shape | Porosity | Time-averaged Total Drag Force (N) | Time-averaged Total Heat Transfer Rate (W) |
|---|---|---|---|---|---|
| A1 | 300 | Circular | 0.50 | 0.00789 | 703.9 |
| A2 | 300 | Circular | 0.87 | 0.00615 | 1049 |
| A3 | 300 | Square | 0.87 | 0.00662 | 1269 |
| A4 | 500 | Square | 0.87 | 0.01607 | 1719 |

Table 4: Summary of simulation cases used in this paper to represent different regimes of flow in porous media.

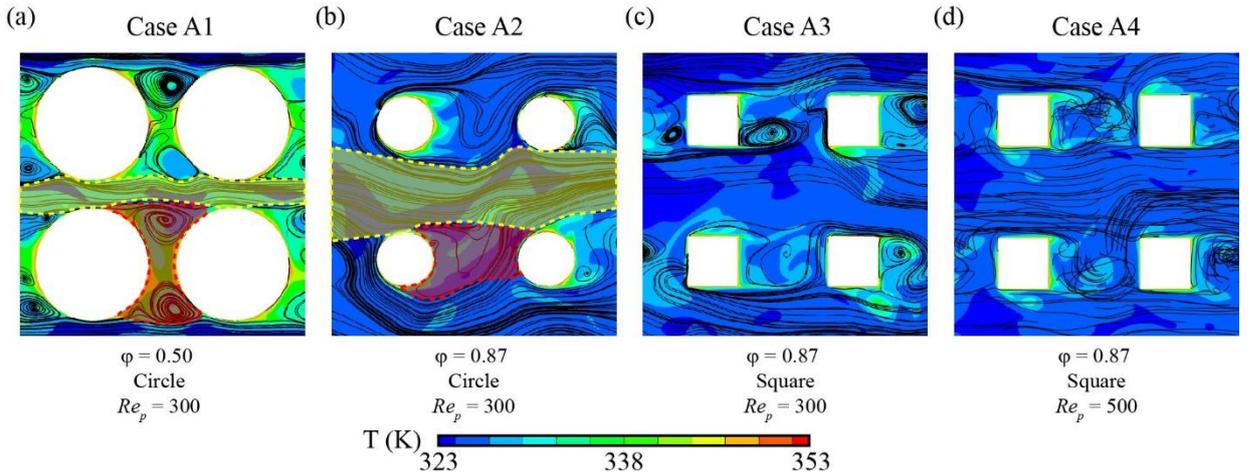

Figure 5: Instantaneous 3D streamlines projected on the *x-y* plane and temperature distribution at $z = 0$ for cases (a) A1, (b) A2, (c) A3, and (d) A4 (table 4). A sub volume of the REV of size ($2s$, $2s$) is shown here. The locations of the streamwise and lateral void spaces are shown in red and yellow dotted lines respectively.

The instantaneous flow field is first examined to identify the influence of the micro-vortices on the temperature distribution inside the porous medium. Qualitative observations are made using the representative cases shown in table 4 for different flow regimes observed in porous media turbulence. The time-averaged total drag force and total heat transfer rate on all of the solid obstacles inside the REV for each case are reported in table 4. The solid obstacle hydraulic diameter and the characteristic temperature difference are identical for all the cases. The time-averaged total drag force is higher for the low porosity case A1 (circle, $\varphi = 0.50$, $Re_p = 300$) when compared to the high porosity cases A2 (circle, $\varphi = 0.87$, $Re_p = 300$) and A3 (square, $\varphi = 0.87$, $Re_p = 300$) due to the constricting geometry in case A1. The time-averaged total heat transfer rate is lower for the low porosity case A1 when compared to the high porosity cases A2 and A3. This indicates that a higher surface area per unit volume is not necessarily favorable since the heat transfer rate is less even though there is a higher drag. At the higher porosity, the time-averaged total drag force for the square cylinder solid obstacles is higher than that of the circular cylinder solid obstacles. However, the time-averaged total heat transfer rate is also higher for the square cylinder solid obstacles compared to that of the circular cylinder solid obstacles. The heat transfer rate is the highest for case A4 (square, $\varphi = 0.87$, $Re_p = 500$) since it has the highest Reynolds number. However, it should be noted that the increased heat transfer comes at the cost of high drag force. These trends in the time-averaged flow can be attributed to the micro-vortices and their dynamics as shown below. For ease of reference, we are dividing the REV into primary and secondary flow regions based on the underlying flow features. In all the cases, we define the primary flow region as the region in between the separation shear layers formed behind the solid obstacle. The primary flow is characterized by virtually unobstructed, high velocity flow from the inlet of the periodic domain. The secondary flow region is the region behind the solid obstacles occupied by the vortices. We define the lateral and streamwise void spaces in the porous medium geometry in the regions of the primary and secondary flows respectively (figures 5 (a) and (b)).

In case A1, recirculating vortex systems are observed in the secondary flow region (figure 5(a)). Sharp gradients in the temperature distribution are present at the boundary between the secondary and primary flow regions. The velocity of the recirculating vortex core is close to zero (stationary). As a result, the vortex core temperature at a statistically steady state is the closest to the solid obstacle surface temperature among all the cases studied here. A high vortex core temperature lowers the temperature gradient at the solid obstacle surface covered by the recirculating vortex. By this mechanism, the recirculating vortex system renders the vortex-covered portion of the solid obstacle surface less conducive for heat transfer, storing heat in the streamwise pore space. In contrast, a more diffuse temperature distribution is observed in the case of a shedding vortex system at high porosity, such as in cases A2-A4. The shedding process carries heat away from the solid obstacle surface that is stored in the vortices. In figures 5(b)-(d), vortex structures with elevated core temperatures are being transported away from the solid obstacle surface. The temperature distribution near the solid obstacle surface in the secondary flow region in cases A2-A4 will have sharper gradients than in case A1. The core temperature of the vortices is also lower in cases A2-A4 than in case A1 since the vortices are not stationary in the streamwise void space. Therefore, the vortex regions in cases A2-A4 have a higher surface heat transfer rate when compared to case A1. The shedding vortex system facilitates convection heat transfer better than the recirculating vortex system.

The change in the solid obstacle shapes between cases A2 and A3 influences the location of the flow separation. For square solid obstacles, the flow separation is prescribed at the sharp corners of the square solid obstacle shape. This results in a smaller intensity of oscillation of the path flowed by the streamlines in the secondary flow region, which is reflected in the time series of the solid obstacle surface forces shown in section 3.2. There is also a consistent contact area between the vortices and the solid obstacle surface throughout the vortex shedding cycle. The change in the Reynolds number between cases A3 and A4 does not appear to change the flow patterns in any apparent way. It is shown in table 4 that both the heat transfer rate and drag force are substantially higher for the high Reynolds number case, case A4, compared to the remaining cases. This increase is not caused by any apparent qualitative change that occurs between cases A3 and A4 when the Reynolds number is increased.

In all of the cases, the micro-vortices introduce spatial inhomogeneity in the heat transfer characteristics on the surface of the solid obstacle. The temporal evolution of the vortex systems in the high and low porosities possesses unique characteristics that determine the heat transfer rate as well. For all the cases, the vortex shedding process starts by vortex formation on the solid obstacle surface due to flow separation in the region of the porous medium with intrinsically adverse pressure gradient. The formation of the vortex entrains cold fluid into the streamwise void space from the primary flow region. The definition of the void space labels is illustrated in figure 5. After the vortex is formed, the vortex core diameter increases while being attached to the surface and the vortex core temperature approaches the solid obstacle surface temperature. Due to the high core temperature, the vortex acts as an "insulation" on the solid obstacle surface in this stage. Next, the vortex detaches from the solid obstacle surface and a new vortex forms as this cycle continues. The evolution of the detached vortex is different for each case. The vortex evolution process is illustrated in figure 6 by plotting the instantaneous flow streamlines. The animated sequence of streamline plots is available in supplementary movies 1-4 for cases A1-A4. The 3D coherent turbulent structures are visualized using iso-surfaces of the Q-criterion in figure 7. The animated sequence of coherent turbulent structures is available in supplementary movies 5-8 for cases A1-A4.

In case A1, the detached vortex (vortex A in figure 6(a-i)) recirculates within the streamwise void space for a shorter period of time than case A2 and A3. The short-lived nature of the vortex in case A1 is best visualized in the animation of the 3D coherent structures in supplementary movie 5. The detached vortex A is visible in figure 7(a) as a tubular coherent structure oriented in the *z*- direction. The tubular structure has deformations and a non-uniform size along the *z*- direction, but consistently appears at every *xy*- plane along the *z*- axis in the streamline plots. The coherent turbulent structures in case A1 are concentrated in the primary flow region and at the lower side of the separation shear layer at the boundary with the secondary flow region. There are very few coherent structures visible in the secondary flow region since the vorticity in the secondary flow region is low. Slow, sustained recirculation over several vortex cycles is responsible for the high temperature inside the secondary flow region. Heat transfer is further diminished by the fact that the micro-vortices are in contact with both solid obstacles surfaces in the streamwise void space. When a new vortex (vortex B in figures 6(a-i) and 7(a)) begins to form and grow in size over time, it is limited in space to the separation shear layer. Vortex B impinges on vortex A as it grows due to the small void space at $\varphi = 0.50$ causing vortex A to deform (figure 6(a-ii)). The instantaneous streamlines then indicate that there is only one vortex that remains, as seen in the bottom half of the streamwise void space in figure 6(a-iii). The outcome of the interaction between vortices A and B is the sustenance of the recirculating motion in the secondary flow region leading to high core temperature (figure 6(a-iii)). This is inferred from both the temperature contours in figure 6(a), and the 3D coherent structures in figure 7 and movie 5. A detailed analysis of the interaction between vortices A and B may be obtained by using a very high grid resolution, but it is not critical for the message in this paper. There is some evidence in the present work that the newly formed vortex B primarily influences the separation shear layer. The temperature distribution at the separation shear layer shows "waves" of high temperature synchronized with the formation of vortex B. The "waves" appear only in the separation shear layer and not inside the streamwise void space. The 3D coherent structures clearly show the formation of vortex B, which is then advected in the separation shear layer. Vortex A consistently recirculates in the streamwise void space until it diminishes and is advected into the primary flow (top half of the streamwise void space in figure 6(a-iii)). The breakup of vortex A is less frequent than the formation of vortex B and the vortex breakup is not periodic in time.

The entire vortex formation process occurs simultaneously on both the upper and lower halves of the streamwise void space, which is different from the alternating characteristic of the von Karman vortex shedding process. The vortex structures are produced by the shear layer between the primary and secondary flow regions due to the Kelvin-Helmholtz (K-H) instability. The K-H instability is not translated into a von Karman instability in case A1 due to: (1) the absence of interaction between the top and bottom shear layers, (2) the absence of the alternating shedding process, and (3) the sustained recirculation in the streamwise void space resembling a lid driven cavity flow at steady state. Vortices A and B are the primary vortices in case A1 on the basis of vorticity magnitude and the relevance to transport in porous media. It is shown in section 3.3 that the interaction between the shear layer and the vortex pair results in a peak in the Nusselt number. This supports the experimental observations of Nguyen *et al.* (2019) that the shear layer and the vortex pair are important coherent structures in porous media turbulence.

The recirculating motion of vortices A and B promote mixing and heat transfer when compared to a still fluid. This can be visualized in the temperature distribution shown in figure 6(a), where the recirculating vortex entrains low temperature fluid into the streamwise void space near the stagnation point. The entrained fluid absorbs heat from the solid obstacle surface and recirculates with a higher temperature. This enhances the heat transfer of the front half of the solid obstacle surface more than the rear half. This is discussed further in section 3.3. Another pair of recirculating vortices is also present in the space between the primary vortices (figures 6(a-i) and (a-ii)), which are the secondary vortices in this case. The secondary vortices are detrimental to convection heat transfer in porous media since they are characterized by low vorticity and their core is virtually stationary. Additionally, the secondary vortices are not in direct contact with the incoming flow (the primary flow) since they are sandwiched between the two primary vortices. Therefore, the temperature in the region occupied by the secondary vortex is higher than that in the region occupied by the primary vortex and the primary flow region. The temperature gradient at the solid obstacle surface in the region occupied by the secondary vortex is low, which results in a low Nusselt number in the part of the solid obstacle surface in this region.

These stages are also shown as an animated sequence of the coherent turbulent structures using the Q-criterion in the supplementary movie 5. The animated sequence shows that a majority of the fast-moving coherent structures are located in the primary flow region. These coherent structures only have a small interaction with the recirculating vortex system (vortex A in figure 7(a)). The turbulent structures in the primary flow have a significantly faster time scale than in the secondary flow. This indicates that a majority of the fast turbulent structures do not come in contact with the solid obstacle surface to effectively engage in convection heat transfer.

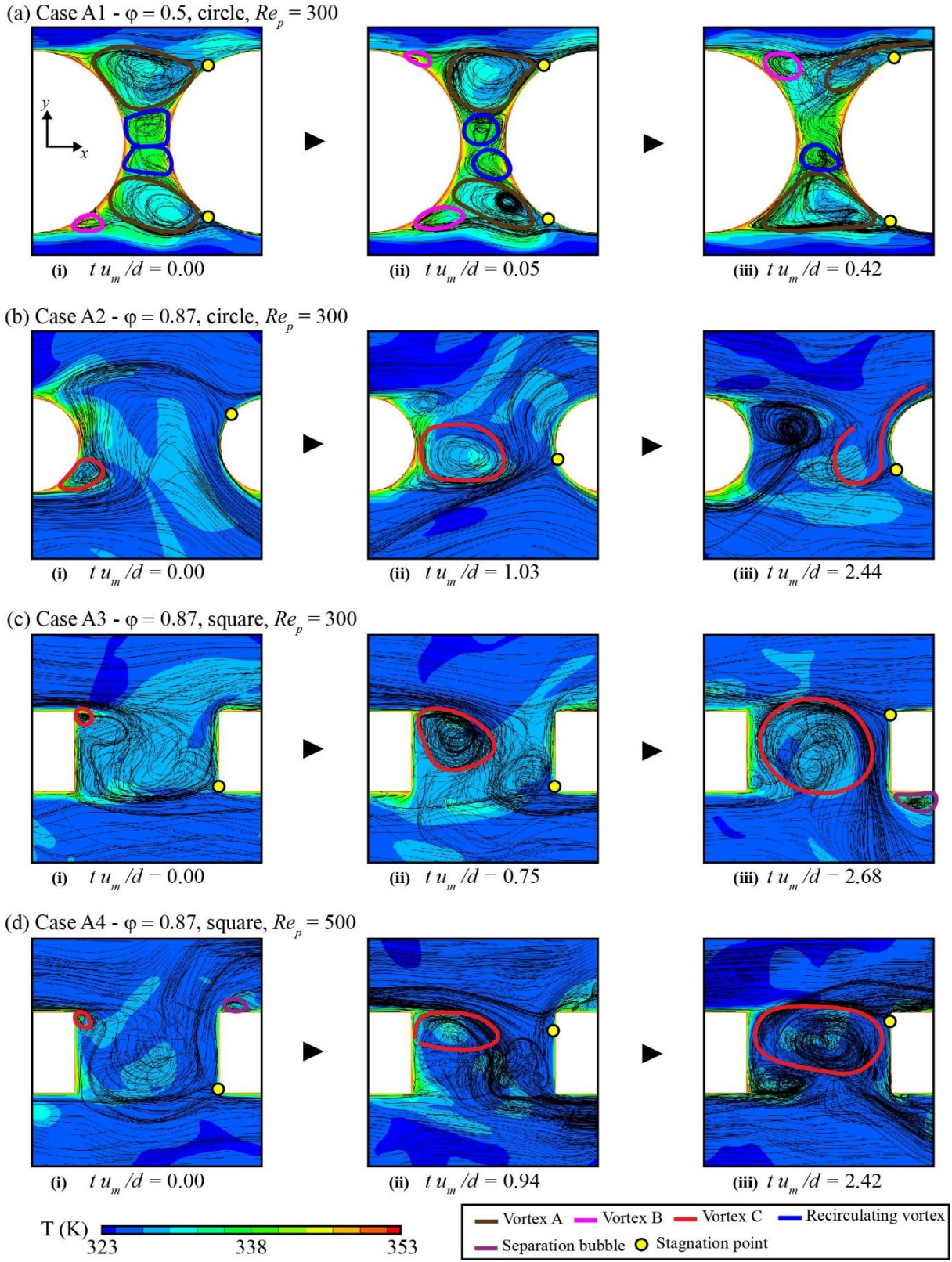

Figure 6: Instantaneous flow streamlines showing the vortex shedding process and its contribution to heat transfer for cases A1-4. Streamlines are 3D streamlines projected onto the *x-y* plane for the solid obstacle in the second row and second column of the REV. Animations of these sequences are shown in supplementary movies 1-4.

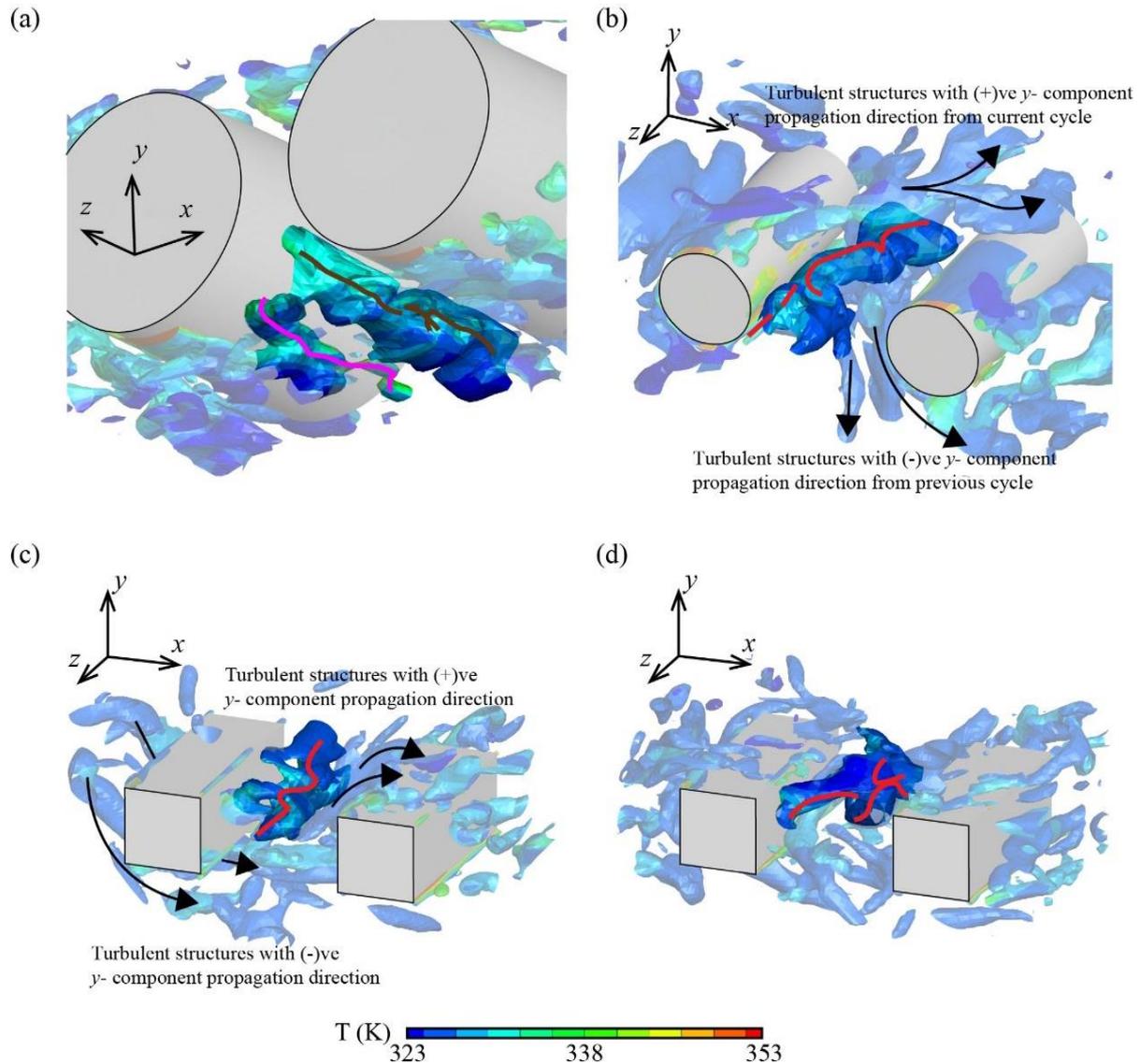

Figure 7: 3D coherent structures visualized using the Q-criterion for cases (a) A1, (b) A2, (c) A3, and (d) A4. Iso-surfaces of the Q-criterion at 0.020 $Q_{max}$ are plotted in this figure. Vortices A, B, and C from figure 6 are highlighted and their vortex core lines are shown as solid lines of colors brown, magenta, and red, respectively. The remaining vortices are shown with 50% color saturation to remove clutter in the figure. The animated sequences are available in supplementary movies 5-8.

In case A2, the porosity is higher than in case A1. The porous medium geometry is less confined since the solid obstacles are farther apart. The vortex shedding process more closely resembles the von Karman vortex shedding in flows around a single cylinder. Vortex formation alternates between the upper side and the lower side of the solid obstacle surface (figure 6(b)). This is different from case A1 where the two vortices on both sides of the solid obstacle form independently. Once the newly formed vortex grows and detaches from the solid obstacle surface, it is advected by the primary flow (vortex C in figure 6(b)). The vortex in case A2 is in contact with only one solid obstacle surface at a given time. The vortex is either attached to the solid obstacle where it is formed or it is impinging on the downstream solid obstacle. Vortex C breaks up after impingement on the downstream solid obstacle and enters the primary flow region (figure 6(b-iii)), where it ceases to recirculate in the *x-y* plane and diminishes in strength. The breakup of vortex C is also influenced by the formation and growth of a new vortex behind the solid obstacle (figure 6(b-ii) and (b-iii)). The vortex C is deformed into coherent turbulent structures elongated in the streamwise direction (figure 7(b)).

Note that the location of the stagnation point on the downstream solid obstacle (yellow dots in figure 6) switches from the upper half of the solid obstacle surface in figure 6(b-i) to the lower half in figure 6(b-ii). This can also be observed in supplementary movie 2. The switch is an indication of a secondary flow instability, which can also be visualized in the tortuosity in the path of the streamlines in figure 6(b). The vortex shedding process in the constrained space of the porous medium in case A2 introduces the secondary instability that causes the fluctuation of the flow separation and stagnation points on the solid obstacle surface. The 3D coherent structures also indicate the presence of the instability, which becomes evident from the propagation direction of the structures. In figure 7(b), the propagation direction of the coherent turbulent structure has a positive *y*- component for the current cycle of the secondary instability and a negative *y*- component for the previous cycle. The frequency of the secondary flow instability is smaller than the frequency of the vortex formation and it is implicitly linked to the vortex shedding process. The secondary flow instability in case A2 is clearly exhibited in the flow visualization since the porosity is high and the solid obstacle shape is circular, making it the least constrained geometry among all the cases studied.

In case A3, the process of vortex formation and shedding is similar to that of case A2. Unlike case A2, the locations of the separation points do not change over time due to the sharp corners of the square obstacle shape (figure 6(c) and movie 3). This confines the vortex formation to only the rear face of the square solid obstacle resulting in a smaller amplitude of the secondary flow instability observed in case A2. Comparing figures 6(b) and (c), there are several features in the flow patterns that are similar between cases A2 and A3. In both cases, the vortex formation is alternating in nature. The shedding vortex interacts with the primary flow, where it is deformed into elongated structures oriented in the streamwise direction (figure 7). This process repeats itself over time. In case A3, a transient separation bubble forms on the surface of the solid obstacle near the vertices on the front face of the solid obstacle due to the tortuosity of the streamline path (figure 6(c-iii) and supplementary movie 3). The location of the separation bubble alternates between the top and bottom surfaces in response to the vortex shedding cycle. Its influence on heat transfer is limited due to its short life span. The tortuous path of the streamlines and the alternation of the location of the stagnation point suggest that the secondary flow instability is present in case A3, similar to case A2. The intensity of the fluctuations in the locations of the flow separation and stagnation points are limited by the sharp corners in the square geometry in case A3 when compared to the smooth circular geometry in case A2. The presence of the secondary flow instability in case A3 can also be verified from the alternation of the propagation direction of the coherent structures in movie 7.

Comparing figures 6(c) and (d), the flow features in case A3 are almost identical to that of case A4 even though the Reynolds number is different. The coherent turbulent structures also have similar shapes and sizes between cases A3 and A4 (figure 7(c) and (d)). The key difference that is brought by increasing the Reynolds number of the flow from case A3 to A4 is the faster time scales in case A4 compared to A3 (movies 7 and 8). Therefore, the high Reynolds number case A4 verifies that the qualitative observations made in this work are independent of the change in the Reynolds number as long as the flow regime does not change.

So far, we have noted the similarities and differences between the vortex systems encountered in cases A1-A4. The following key observations are used throughout the remainder of the discussion. The main differences in vortex transport are as follows. In case A1, the vortices remain localized within the streamwise void space. In cases A2 and A3, the vortices leave the streamwise void space and enter the lateral void space as elongated turbulent structures. This difference is brought about by the distance between the solid obstacle surfaces due to the different porosities. Cases A3 and A4 have identical flow features indicating that the observations will be valid after a change in the Reynolds number within the same regime of turbulent flow. The recirculating micro-vortex core in case A1 is stationary, but the surrounding fluid has some rotational velocity that contributes to heat transfer. The shedding vortices are more effective in enhancing heat transfer due to two momentum transport processes: micro-vortex advection from the surface and turbulent mixing with the primary flow. Heat transfer due to micro-vortex advection is observed to be sensitive to the geometry of the porous medium. At this stage we conclude that convection heat transfer is primarily affected by the following:

1. The solid obstacle surface area in contact with the recirculating vortices.
2. The length and time scales of the shedding vortices.
3. The dynamics of micro-scale flow instabilities.

*3.2. Comparison of the Dynamics of the Surface-Averaged Heat Transfer and the Drag Force*

To simplify the analysis of the flow instabilities, the three-dimensional microscale flow field is reduced to a macroscale level. The macroscale momentum conservation equation defined in equation 3.1 (according to the work of de Lemos 2012) is used as the basis to discover the significant terms. The operator $\langle - \rangle^i$ indicates volume average in the fluid domain. The volume of the REV is denoted by $\Delta V$. The interfacial area between the solid and fluid phases in the REV is denoted by $A_{interface}$ and $n_i$ is the normal vector of $A_{interface}$. For all the cases, the pressure drag in the momentum budget has a higher magnitude than the viscous drag and is therefore more dominant (shown for Case A1 in figure 8(a)). Since the intensity of the oscillation of the pressure drag is similar to that of the inertial component, we conclude that the dynamics of the flow can be analyzed by looking at the pressure drag component. However, the total drag force is used in the analysis since the viscous drag component has a negligible influence on the dynamics of the pressure drag. The phase difference between the vortex motions behind the individual solid obstacles is an important consideration in this analysis. Interference is expected when summing the forces acting on all of the solid obstacles in the REV. To eliminate the influence of the phase difference, the total surface forces $F_i$ acting on a single solid obstacle and its pressure ($F_{p,i}$) and viscous ($F_{v,i}$) components are used in the subsequent analysis. The details of the phase difference and its influence on the macroscale force terms are shown in Appendix C. The forces are standardized for analysis as shown for the pressure drag force ($F_{p,1}$) in equation (3.2). The asterisk notation in equation (3.2) indicates the standardization operation. $N$ in equation (3.2) denotes the number of time steps in the signal.

$$\underbrace{\rho \frac{\partial}{\partial t}\left(\varphi \langle u_i \rangle^i\right)}_{inertial} = \underbrace{\rho \varphi g_i}_{applied} + \underbrace{\frac{\mu}{\Delta V} \int_{A_{interface}} n_j \, \partial_j u_i \, dS}_{viscous\ drag} - \underbrace{\frac{1}{\Delta V} \int_{A_{interface}} n_i p \, dS}_{pressure\ drag} \qquad (3.1)$$

$$F^*_{p,1} = (F_{p,1} - \overline{F_{p,1}}) / \sqrt{\sum (F_{p,1} - \overline{F_{p,1}})^2 / (N-1)} \qquad (3.2)$$

The oscillation of the standardized pressure drag ($F^*_{p,1}$) with time (figure 8(b)) will consist of two important instabilities in all the cases as explained in section 3.1. There exists a small time scale oscillation in the standardized pressure drag that is caused by the vortex shedding process. There is also a large time scale oscillation caused by the secondary flow instability. These instabilities are not readily visible in the plots of $F^*_{p,1}$ with time due to the presence of randomness in turbulence and the interference of the vortex motions at the neighboring solid obstacles. This suggests the need for the frequency analysis that is presented in figure 9. The contrast in the dynamics between the low and high porosity cases is clearly visible in figure 8(b). The oscillations in the low porosity case A1 (circle, $\varphi = 0.50$, $Re_p = 300$) are more rapid than those of the high porosity cases ($\varphi = 0.87$). This is due to the presence of the K-H instability in case A1 instead of the von Karman instability seen in cases A2-A4. The large magnitude of streamwise velocity in the primary flow region in case A1 is also a contributing factor. The secondary flow instabilities are not explicitly visible in figure 8(b). The secondary flow instability is caused by the stagnation of the flow on the neighboring solid obstacle due to the vortex impingement on the solid obstacle surface. In case A1, the secondary flow instability was not explicitly visible in the flow streamlines in section 3.1. It appears in figure 8(b) in the form of the minima and the maxima in the pressure force magnitude. The amplitude of oscillation of $F^*_{p,1}$ is larger at $tu_m/d = 25$ than at $tu_m/d = 10$ (figure 8(b)). In case A1, the adverse pressure gradient introduced by the stagnation pressure in the converging geometry of the porous medium competes with the inertial response of the flow (see pressure and inertial forces in figure 8(a)). In cases A2-A4, the secondary flow instability is visible in the flow visualization as shown in section 3.1. In this case, the stagnation pressure sways the direction in which the advected vortices travel. The increased space in between the solid obstacles in cases A2-A4 is responsible for the different mechanism of the secondary flow instability when compared to case A1. The formation of the alternating vortices introduces only one stagnation point in cases A2-A4, when compared to two stagnation points in case A1. The secondary flow instability in cases A2-A4 can only be seen in the frequency spectra shown in figure 9.

Since the dynamics of the vortices is characterized by multiple frequencies, the Fast Fourier Transform (FFT) technique is used to segregate the time series signal into different frequencies. It should be noted that the FFT technique that is used in this work assumes that the input is periodic. Since turbulent flows do not periodically repeat, periodicity is artificially imposed by mirroring the input. To determine the relationship between the heat transfer dynamics and the flow instabilities, the surface averaged Nusselt number ($Nu_m$), the Coefficient of Drag ($C_D$), and the Coefficient of Lift ($C_L$) are analyzed. The Nusselt number ($Nu$) is calculated as per equation (3.3), where $q_w$ is the heat flux at the solid obstacle surface and $T_{in}$ is the bulk temperature at the inlet plane of a column of solid obstacles in the REV. The distribution of $Nu$ is then averaged over the surface area of a single solid obstacle to determine $Nu_m$. The conventional definition of the characteristic temperature difference that uses the free stream temperature to calculate the Nusselt number is not suitable for periodic porous media. In the present simulations, the bulk temperature at the inlet of the REV is identical across all the cases. However, the bulk temperature will increase in the streamwise direction depending upon the heat transfer characteristics of the porous medium geometry. The bulk temperature at the inlet plane of each column of solid obstacles in the REV is taken as one of the reference temperatures to remove the upstream dependence of the Nusselt number. This method provides a more consistent value of $Nu_m$ that is similar for all the solid obstacles in the REV. The Coefficient of Drag ($C_D$) and Coefficient of Lift ($C_L$) are calculated as per equations (3.4) and (3.5), where $S$ is the surface area of a single solid obstacle and $F_i$ is the total force on a single solid obstacle. Note that the ratio of $A_{interface}$ to $S$ gives the number of cylinders in the REV. All of the signals are standardized as per the procedure given in equation (3.2). Standardizing the signal scales it to have a mean of zero and a standard deviation of 1 to allow a direct comparison of the dynamics of the surface forces and heat transfer.

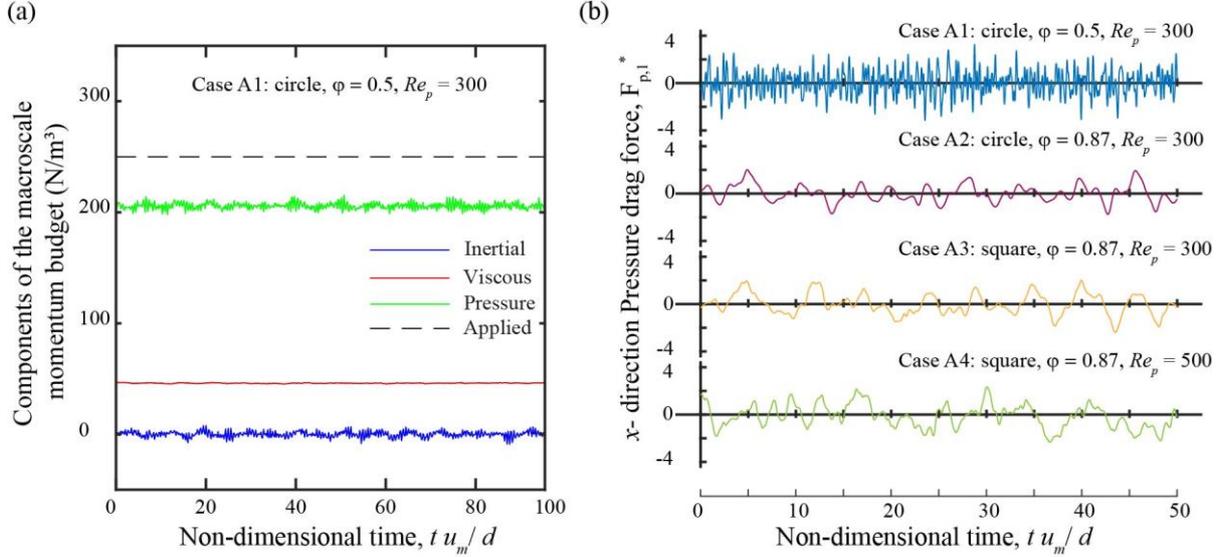

Figure 8: (a) Momentum budget of forces in the *x*-direction within a REV for a flow in a porous medium composed of circular cylinders for case A1. The time averaged error in the conservation of macroscale momentum is 0.73%. (b) Standardized *x*- direction pressure drag force acting on a single solid obstacle in row 2, column 2 of the REV.

$$Nu = \frac{q_w d}{(T_w - T_{in})k} \quad (3.3)$$

$$C_D = \frac{F_1}{0.5\rho u_m^2 S} \quad (3.4)$$

$$C_L = \frac{F_2}{0.5\rho u_m^2 S} \quad (3.5)$$

The frequency spectra of $C_D^*$ and $C_L^*$ are used to identify the features of the dynamics of the surface flow phenomena. The frequency spectrum of $Nu_m^*$ is then correlated with the observations about the surface flow phenomena to understand their influence on heat transfer. The frequency scale is non-dimensionalized using $u_m$ and $d$ to yield the non-dimensional frequency *f*. The non-dimensional frequency has a similar form to the Strouhal number, which is typically used to report the vortex shedding frequency. We use the term non-dimensional frequency in this work since there are other instabilities present in the flow. The absolute value of the Fourier coefficients is used to indicate the contribution of the individual frequencies in the signal. Note that the absolute value of the Fourier coefficient in the frequency domain does not directly correspond to the intensity of the oscillations in the time domain. Note that the power of the oscillations of the signals above the frequency of ~$10^2$ are more than two orders of magnitudes smaller than the time averaged value. This indicates that the influence of the smaller eddies is insignificant in this study. The frequency range above $10^2$ has been verified to be numerical noise in all the cases using the following procedure. The coefficients of the amplitude spectrum of the signal above the frequency of $10^2$ is set equal to zero. The modified signal is transformed back to the time domain and compared to the original signal. The curve comparison is shown in figure S2 in the supplementary material.

In case A1 (circle, $\varphi = 0.50$, $Re_p = 300$), two distinct peaks in the $C_D^*$ and $C_L^*$ spectra are observed at $f = 1.1$ and $3.6$ (figure 9(a)). Note that there is a greater amount of noise in the spectrum for $C_D^*$ compared to $C_L^*$ that arises from randomness in the oscillations. The absolute value of the Fourier coefficients is the highest at $f = 3.6$, which is identical to the frequency of vortex formation obtained from flow visualization (Movie 1). The peak at the lower frequency of $f = 1.1$ corresponds to a non-stationary oscillation that is introduced by the secondary flow instability. The secondary flow instability is not as periodic as the vortex shedding process and is therefore characterized by a wider base width at the location of the peak in the spectrum in figure 9(a). When the contribution of all the frequencies that constitute the peak in the spectrum at $f = 1.1$ are added, the large time scale oscillation corresponding to the secondary flow instability reported in figure 8(b) is recovered. The frequency spectrum of $Nu_m^*$ shares the dominant peak at $f = 3.6$ with the spectra of $C_D^*$ and $C_L^*$. This suggests that the dynamics of heat transfer is governed in part by the dynamics of the vortex shedding for case A1. There is a minor peak in the frequency spectrum of $Nu_m^*$ at $f = 1.1$. However, its significance is far too small to establish a direct correlation with the secondary flow instability. The remainder of the frequency spectrum of $Nu_m^*$ yields a large time scale oscillation that is a secondary instability in heat transfer. There is no apparent coincidence between the secondary flow and thermal instabilities in case A1.

In case A2 (circle, $\varphi = 0.87$, $Re_p = 300$), the frequency spectrum shifts to a lower frequency range that is one order of magnitude less than that in case A1. One of the reasons for the difference is the velocity of the flow in the primary flow region, which is higher for case A1 due to the low porosity. Another reason is that the vortex systems are entirely different between low porosity (A1) and high porosity (A2) cases as discussed in section 3.1. Note that this is the case for all high porosity cases (A2-A4) compared to the low porosity case (A1). There are 4 significant frequencies observed in the spectra of $C_D^*$, $C_L^*$, and $Nu_m^*$ for case A2 (figure 9(b)). The significance of each frequency varies depending on the signal. Note that there is less noise in the spectra for case A2 compared to case A1. The peak in the frequency spectrum at $f = 0.17$ is similar for $C_D^*$, $C_L^*$, and $Nu_m^*$ and has the largest contribution to the signals. The two smaller peaks in the frequency spectrum of $C_L^*$ at $f = 0.27$ and $0.35$ are less prominent in the frequency spectrum of $Nu_m^*$. From the flow visualization, the frequency of vortex shedding on each side of the solid obstacle is estimated to be $0.3$. The corresponding peaks in the frequency spectrum are $f = 0.27$ and $0.35$. The vortex shedding process is split between two peaks in the spectrum due to the inherent randomness in the flow. The frequency estimate of the secondary flow instability from flow visualization is $0.15$, which corresponds well with the peak in the spectrum at $f = 0.17$. The large magnitude of the Fourier coefficients at $f = 0.17$ suggests that the oscillations introduced by the secondary flow instability are dominant. The large amplitude of oscillation introduced in the flow as a result of the secondary flow instability follows from the discussion in section 3.1. There is a large peak in the frequency spectrum of $Nu_m^*$ at $f = 0.1$ that is also visible in the frequency spectrum of $C_D^*$. The correlation between these peaks and the flow features is not apparent and may require non-linear modal analyses to unravel.

Cases A3 and A4 (square, $\varphi = 0.87$, $Re_p = 300$ and 500) exhibit a distinct behavior where the frequency spectra of $C_L^*$ bear no resemblance to those of $C_D^*$ and $Nu_m^*$. Both cases A3 and A4 use square solid obstacles in the GPM that do not have the axisymmetry that circular obstacles possess. There is a strong anisotropy in the dynamics of the forces that act on the surface of square solid obstacles. For the square obstacles, only the horizontal surfaces that are in contact with the primary flow contribute towards the lift force on the solid obstacle. The vertical surfaces that are in contact with the vortices in the secondary flow contribute towards the drag force on the solid obstacle. Note that the spectrum of $C_L^*$ indirectly reflects the instabilities that are associated with the vortex formation through the shear layers in contact with the horizontal surface. The primary source of oscillations in the surface force on the vertical surface is the impingement of microvortices. This is one possible reason why the frequency spectrum of the drag force contains random fluctuations. From the flow visualization, the frequency of vortex shedding is estimated as 0.27. This frequency is highlighted in the frequency spectrum of $C_L^*$ since the vortex shedding process causes the stagnation pressure near the vertices of the square geometry to oscillate along with the shear layer. The root mean squared value of the fluctuation of the static pressure on the surface of the square geometry is the highest at the locations of the stagnation point and vortex impingement on the solid obstacle. The heat transfer is enhanced in these locations as shown later in section 3.3. Therefore, there is a greater similarity between the spectra of $C_D^*$ and $Nu_m^*$ when compared to $C_L^*$. This further supports the notion that the microvortices determine the dynamics of heat transfer in porous media. There is not an appreciable difference in the features of the spectra when the Reynolds number is increased from 300 to 500. This suggests that the observations will hold as long as the flow features do not transition to a different regime of turbulent flow in porous media.

For the case of the flow around an isolated long circular cylinder, laminar-turbulent transition effects are present at $Re_p = 300$ (Williamson 1996), which are not observed in the case of the porous medium. The Strouhal number of vortex shedding for the isolated cylinder is ~0.2 (Fey *et al.* 1998). In the case of two-tandem circular cylinders with identical diameters, the Strouhal numbers of vortex shedding are 0.196 and 0.12 at a Reynolds number of 27,200 (Alam & Zhou 2008). Two Strouhal numbers are reported for the flow around two-tandem circular cylinders due to the interaction between the vortex shedding processes behind the two cylinders. In the present work, the Strouhal number of vortex shedding for the porous medium varies with porosity (corresponding to the peak with the maximum amplitude in figure 9). At $\varphi = 0.50$, the Strouhal number of vortex shedding is 3.6, where the Strouhal number is the value of *f* at the vortex shedding frequency. The Strouhal number for the low porosity porous medium is one order of magnitude greater than that of the isolated and tandem cylinders. At $\varphi = 0.87$, the Strouhal number of vortex shedding is 0.3, which is closer to that of the isolated and tandem cylinder cases than at $\varphi = 0.50$. The difference between the high and low porosity cases highlights the contrast in the underlying flow physics with the change in porosity.

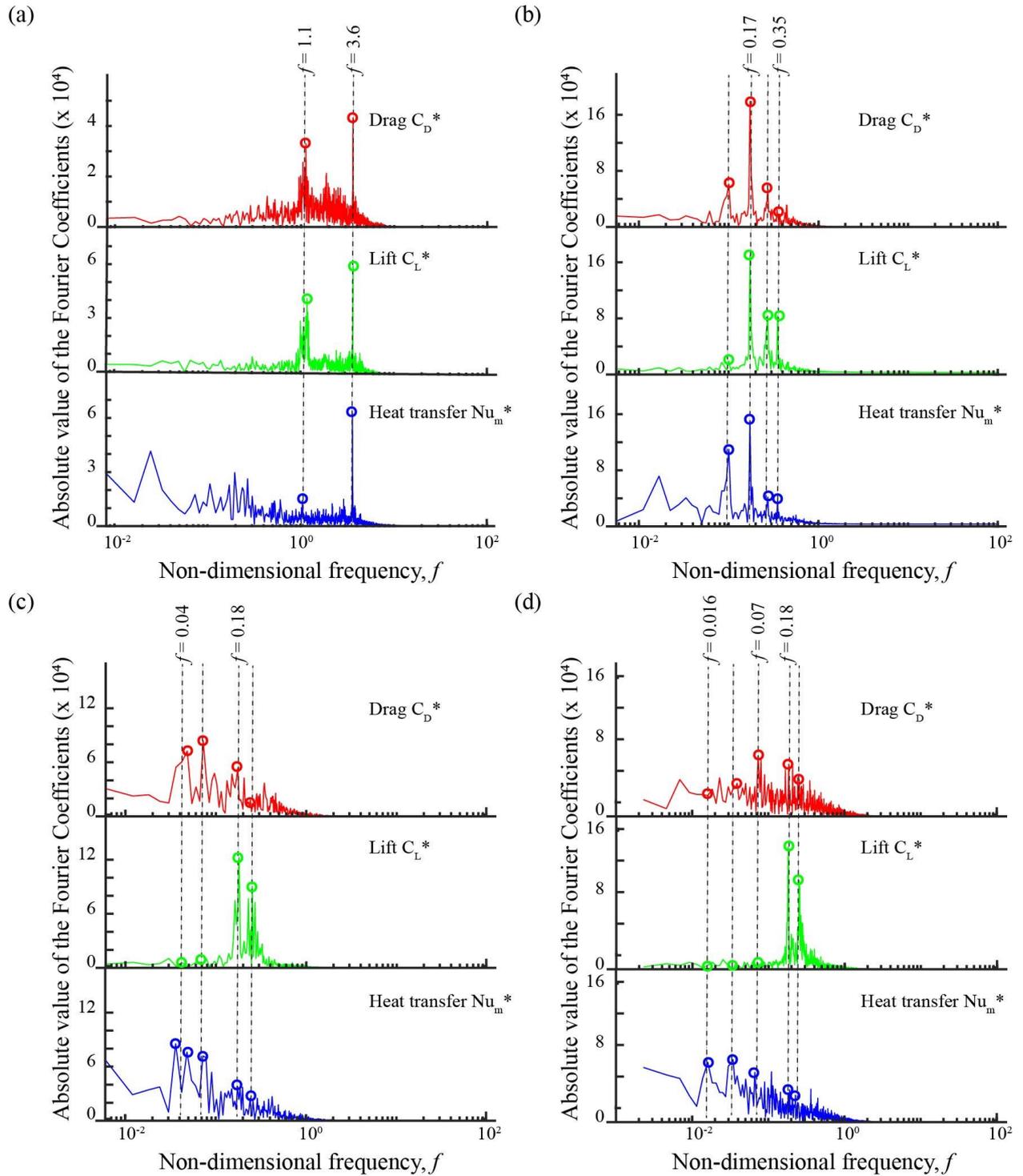

Figure 9: The frequency spectra of the standardized Drag Coefficient ($C_D$), Lift Coefficient ($C_L$), and the surface averaged Nusselt number ($Nu_m$) for cases (a) A1, (b) A2, (c) A3, and (d) A4. The solid obstacle in the second row and second column of the REV is used for the analysis. It is shown in figure S3 in the supplementary material that all of the solid obstacles have similar spectra plots. The vertical axes show the absolute values of the complex Fourier coefficients obtained from FFT.

*3.3. The contribution of the different flow features towards heat transfer*

Following the discussion in section 3.1, the flow features that are typically encountered in flow in porous media are: micro-vortices, flow stagnation, separation shear layer, and secondary flow instabilities. In section 3.2, the prominence of the flow instabilities in heat transfer dynamics has been demonstrated. Next, the spatial distribution of heat transfer on the surface of the solid obstacle and its contribution towards the surface-averaged heat transfer is investigated. The surface Nusselt number distribution on the solid obstacles for cases A1-A4 are plotted in figure 10 along with the surface skin friction lines. The plots are shown for a single instance in time, but the following observations have been verified over multiple vortex system cycles.

Generally, the Nusselt number $Nu$ has a higher magnitude in the vicinity of flow stagnation when compared to the lower magnitude observed in the surface areas with separated flow. Flow stagnation is always associated with vortex impingement in all the cases shown in this paper. Note that the Nusselt number distribution is not uniform in the $z$- direction due to the three-dimensional effects of turbulence. The $z$- component of the skin friction lines on the solid obstacle surface are significant in the regions of the separated flow. The skin friction lines are virtually parallel to the circumferential direction in the remaining regions. Therefore, the location of the vortices on the solid obstacle surface can be identified using the cusps in the skin friction lines. Sharp peaks in the Nusselt number appear over time when a micro-vortex is incident on the solid obstacle surface. This is inferred from the fact that the peaks are always associated with a cusp in the skin friction lines. Note that the orthogonal lines to the skin friction lines are called the vortex lines that provide an indication of the location of swirling motion. Localized peaks in the Nusselt number distribution are observed in the regions of separated flow (yellow bubbles in figure 10). The peaks are observed during vortex formation, where the rotation of the vortex promotes entrainment of fluid from the primary flow. Even though the attached vortices are typically associated with a low Nusselt number, the formation of these localized peaks and the vortex impingement over time is the mechanism by which the shedding vortices promote heat transfer. Vortex dynamics introduces fluid mixing inside the porous medium, which is beneficial for convection heat transfer.

Due to the changes in the flow features, the distribution of Nusselt number on the surface of the solid obstacle varies from case to case. For case A1, there are two banded regions (at $\theta \sim 0.5\pi$ and $\theta \sim 1.5\pi$) with high $Nu$ surrounding the two stagnation lines formed by the recirculating vortex system (figure 10(a)). Banded regions are formed since the vortex structures have a tubular shape elongated in the $z$- direction. High $Nu$ is experienced close to the stagnation line due to the bifurcation of the cold fluid from the primary flow around the stagnation line. The entrainment of the primary flow into the recirculating vortex system at the stagnation line was also observed in section 3.1. Low $Nu$ (<10) is observed on the surfaces in contact with the secondary vortices that are only present in case A1 due to their slow rotation and limited interaction with the primary flow. For case A2, there is a wide band of high $Nu$ that oscillates about the mean position of $\theta \sim \pi$ (figure 10(b)). The stagnation line is also oscillating about that position due to the secondary flow instability. The oscillation of the stagnation line in the case A2 have a higher amplitude than in case A1 due to the increased void space in case A2. This also highlights the distinction in the mechanism of the secondary flow instability between cases A1 and A2. For case A1, there is virtually no positional change in the stagnation line, but the magnitude of stagnation pressure is oscillating. For case A2, both positional and magnitude oscillations are observed. For cases A3 and A4, the high $Nu$ regions are located close to corners 3 and 0 of the solid obstacle (figure 10(c)). This coincides with the location of the stagnation lines at the corners of the square solid obstacle. Since the vortices are in contact with the primary flow region, it enhances mixing and creates regions of elevated $Nu$ outside the stagnation line region.

To determine the fractional contribution of the different flow features towards heat transfer, the joint Probability Density Function (PDF) of the surface Nusselt number and the stress coefficients is calculated. There are 100 bins for each variable yielding a total of 10,000 bins for the entire joint PDF. The suggested sample size for the convergence of the PDF distribution is 150 non-dimensional time units. The PDF distribution is shown here for a single solid obstacle in the second row and second column of the REV. The PDF distribution has been verified to be identical for different solid obstacles in the REV. The histogram for the PDF is segregated based on the flow features that appear in that region on the solid obstacle surface. The boundaries of the regions are determined from the time-averaged flow streamlines. The 'impingement' region is located on the front face of the solid obstacle and is characterized by impinging microvortices and flow stagnation. The boundaries of this region are set at the locations where the local time-averaged static pressure is 50% of the time-averaged stagnation pressure. The 'separated' region is located on the rear face of the solid obstacle and is characterized by separated flow due to vortex formation. The boundaries of the 'separated' region are at the locations of flow separation in the time-averaged flow streamlines. The 'primary' region is in contact with the primary flow. The surface area regions on the solid obstacle that do not belong to the 'impingement' or 'separated' regions form the 'primary' region. The regions are indicated in figure 11. Note that there are sub-regions for the 'impingement' region in case A1 that are not present in cases A2-A4. There are distinct stagnation and recirculation regions due to the unique vortex system at the porosity of 0.5 as discussed in section 3.1. For consistency, the sub-regions are combined in figure 11. There are 3 regions for each case in the PDF plot in figure 11: impingement, separated, primary, and the PDFs corresponding to each of the regions are assigned red, green, and blue color channels, respectively. The PDFs are co-plotted yielding uniquely colored contours that result from overlaying the red, green, and blue color images. Purely red, green, or blue regions in figure 11 imply that only one region contributes to the PDF in that Nusselt number and skin friction coefficient range. Mixed colors indicate that more than one region contributes to the PDF in that Nusselt number and skin friction coefficient range. The individual plots are shown in the supplementary information in figures S4-S11.

Both the pressure and skin friction coefficients are used to calculate the joint PDF with the Nusselt number. The magnitude of shear stress on the surface is used to calculate the skin friction coefficient. The stresses are normalized with $0.5\rho u_m^2$ to obtain the stress coefficients. Pressure drag is the dominant surface force on the solid obstacle surface that has been shown to have a direct correlation with the Nusselt number. Shear stress is also an effective parameter to identify the contribution of the micro-vortices. The micro-vortices are characterized by a low magnitude of shear stress on the surface of the solid obstacle due to the separated flow and the stagnation at the time of impingement. The high shear regions are the boundary layers that are formed in between the solid obstacle surfaces in the transverse direction (normal to the streamwise direction). These regions are typically encountered in the locations where the primary flow directly interacts with the solid obstacle surface. To confirm the observations regarding the flow features, we plotted the locations of the sample points of the PDF that correspond to the chosen criterion on the solid obstacle surface and compared them to the skin friction lines. These plots are for verification purposes only and are not shown in the paper.

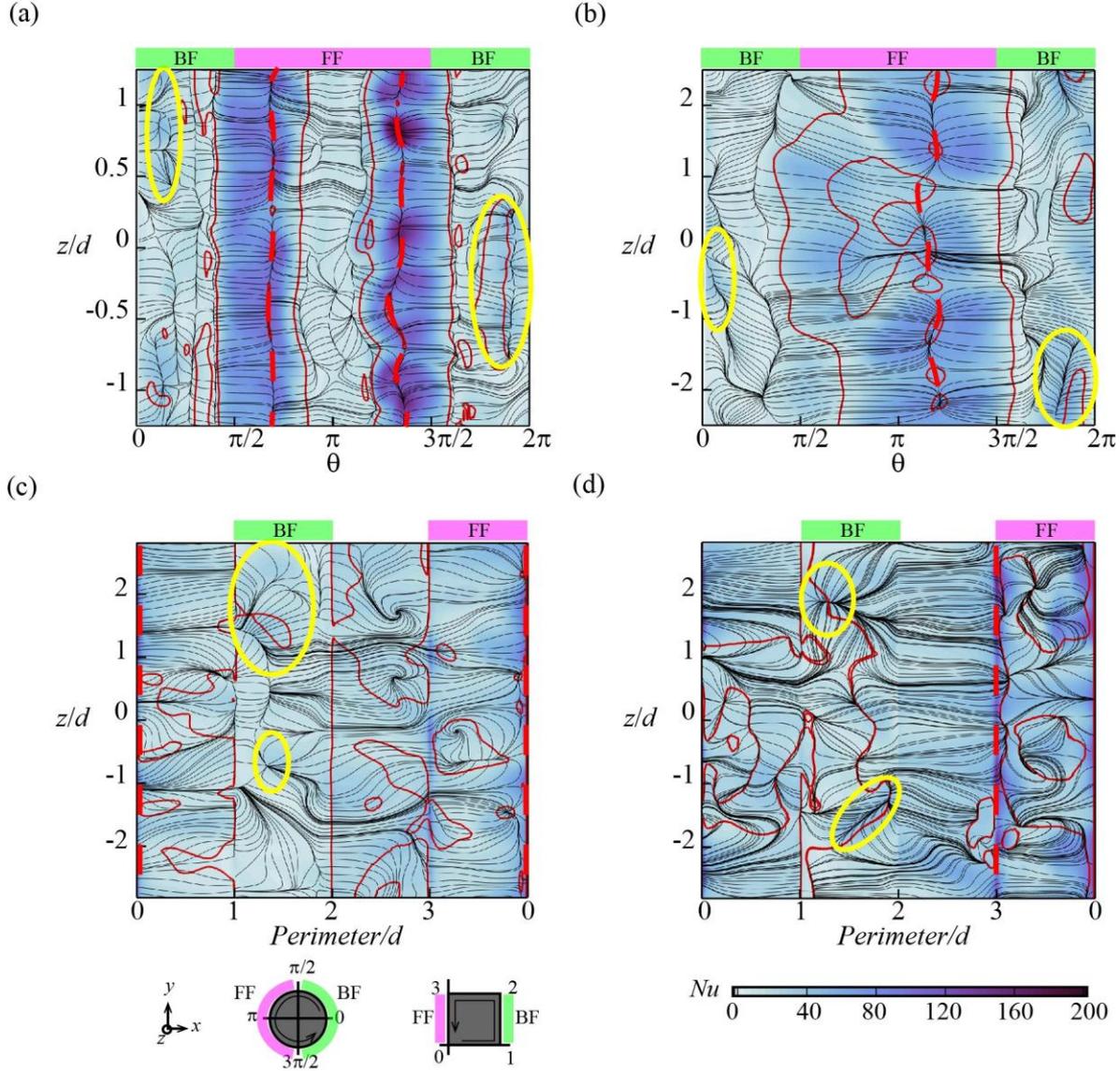

Figure 10: Skin-friction lines (black solid lines) plotted on the surface of the solid obstacles. Iso-lines (red solid lines) of zero shear stress indicate flow separation and the cusp of the skin friction lines indicates the locations of flow stagnation (red dashed lines, referred to as stagnation lines) for cases (a) A1, (b) A2, (c) A3, and (d) A4. Yellow circles indicate the locations of Nusselt number peaks due to vortex formation. The label FF indicates the Forward Facing side and the label BF indicates the Backward Facing side.

In case A1 (circle, $\varphi = 0.50$, $Re_p = 300$), the 'stagnation' sub-region experiences Nusselt number values that are consistently higher than the time and surface averaged Nusselt number. The $C_{pressure}$-$Nu$ and $C_f$-$Nu$ PDF distributions in the 'stagnation' region have the lowest peak probability and the largest variance among all the other regions. This is caused by the intense fluctuations in the flow that is observed in the 'stagnation' region. When the 'stagnation' sub-region is combined with the 'recirculation' sub-region to give the 'impingement' region, the corresponding PDF distribution includes the low $Nu$, $C_{pressure}$, and $C_f$ contributions of the 'recirculation' sub-region (red contour plots in figure 11(a) and figure 12(a)). The 'separated' region exhibits the smallest range of $C_f$, $C_{pressure}$ and $Nu$ values, such that the $C_f$-$Nu$ and $C_{pressure}$-$Nu$ PDF distributions are concentrated near the minimum values of pressure, skin friction, and Nusselt number (green contour plots in figure 11(a) and figure 12(a)). The 'separated' region covers the largest area on the solid obstacle surface, insulating it from the incoming cold fluid. The 'impingement' and 'primary' regions have anisotropic $C_f$-$Nu$ PDF shapes that are clustered along a diagonal line suggesting a strong dependence between shear and heat transfer in these regions. Similarly, the $C_{pressure}$-$Nu$ PDF shapes for the 'impingement' and 'primary' regions are also anisotropic. It can be inferred from this that a majority of the flow dynamics is limited to the 'impingement' and 'primary' regions. A major contributing factor to the high Nusselt number in the 'primary' region is the proximity to the stagnation region. Direct contact with the primary flow is a minor contributing factor. The probability values are biased towards the minimum values of pressure, skin friction, and Nusselt number in the 'impingement' and 'primary' regions, similar to the 'separated' region. Small regions characterized by high skin friction, pressure, and Nusselt number do exist in the 'impingement' and 'primary' regions, unlike in the 'separated' region.

In case A2 (circle, $\varphi = 0.87$, $Re_p = 300$), the vortex system is different from case A1 as shown in section 3.1. The higher porosity also means that there is no clear distinction between the primary and secondary flow regions. The $C_f$-$Nu$ and $C_{pressure}$-$Nu$ PDF shapes are anisotropic and coincident in the 'impingement' and 'primary' regions (figure 11(b) and figure 12(b)). The $C_{pressure}$-$Nu$ PDF is centered at zero gauge pressure in all the regions with an identical $C_{pressure}$ range indicating the strong oscillatory nature of the vortex system in this case and the lack of a clear distinction between primary and secondary flow regions. The shape of the $C_f$-$Nu$ PDF in the 'impingement' region in case A2 combines the features of the 'stagnation' and 'impingement' regions in case A1. The 'separated' region in case A2 contributes high values of $Nu$, unlike in case A1. This feature highlights the presence of shedding vortices in case A2 that were observed in section 3.1 and their importance in promoting heat transfer. The transport of the microvortex away from the surface of the solid obstacle entrains cold fluid into the separated region and promotes heat transfer.

The features of the $C_f$-$Nu$ PDFs in the 'impingement' and 'separated' regions for Cases A3 and A4 (square, $\varphi = 0.87$, $Re_p = 300$ and $500$) are closer to that of case A1 than case A2, even though cases A2-A4 have the same porosity. The square geometry of the solid obstacles in cases A3 and A4 introduces inhomogeneity in the flow that causes the segregation of the primary and secondary flow regions. The microvortices in the square geometry are not transported outside the secondary flow region. Therefore, the flow features appear similar to that of case A1 with additional characteristics that arise from the vortex transport inside the secondary flow region due to the larger pore space. The $C_{pressure}$-$Nu$ PDF shapes in all of the regions for cases A3 and A4 are similar to that of case A2 since the vortex systems are similar, such that the magnitude of pressure dominates the other differences in the flow features. However, the $Nu$ range varies considerably across the cases since large Nusselt number values are observed at the vertices of the square geometry in cases A3 and A4. Since the probability is concentrated at low $Nu$ conditions, the $Nu$ range has a limited effect on the surface averaged $Nu$. The $C_{pressure}$-$Nu$ and $C_f$-$Nu$ PDFs in the 'primary' region are similar to those of case A2 since the primary flow has a stronger dependence on the space in between the solid obstacles than the solid obstacle shape. The PDF shape for the square solid obstacles does not change significantly between the case A3 at $Re_p = 300$ and case A4 at $Re_p = 500$. Similar behavior was observed for the frequency spectrum of the macroscale variables in figure 9.

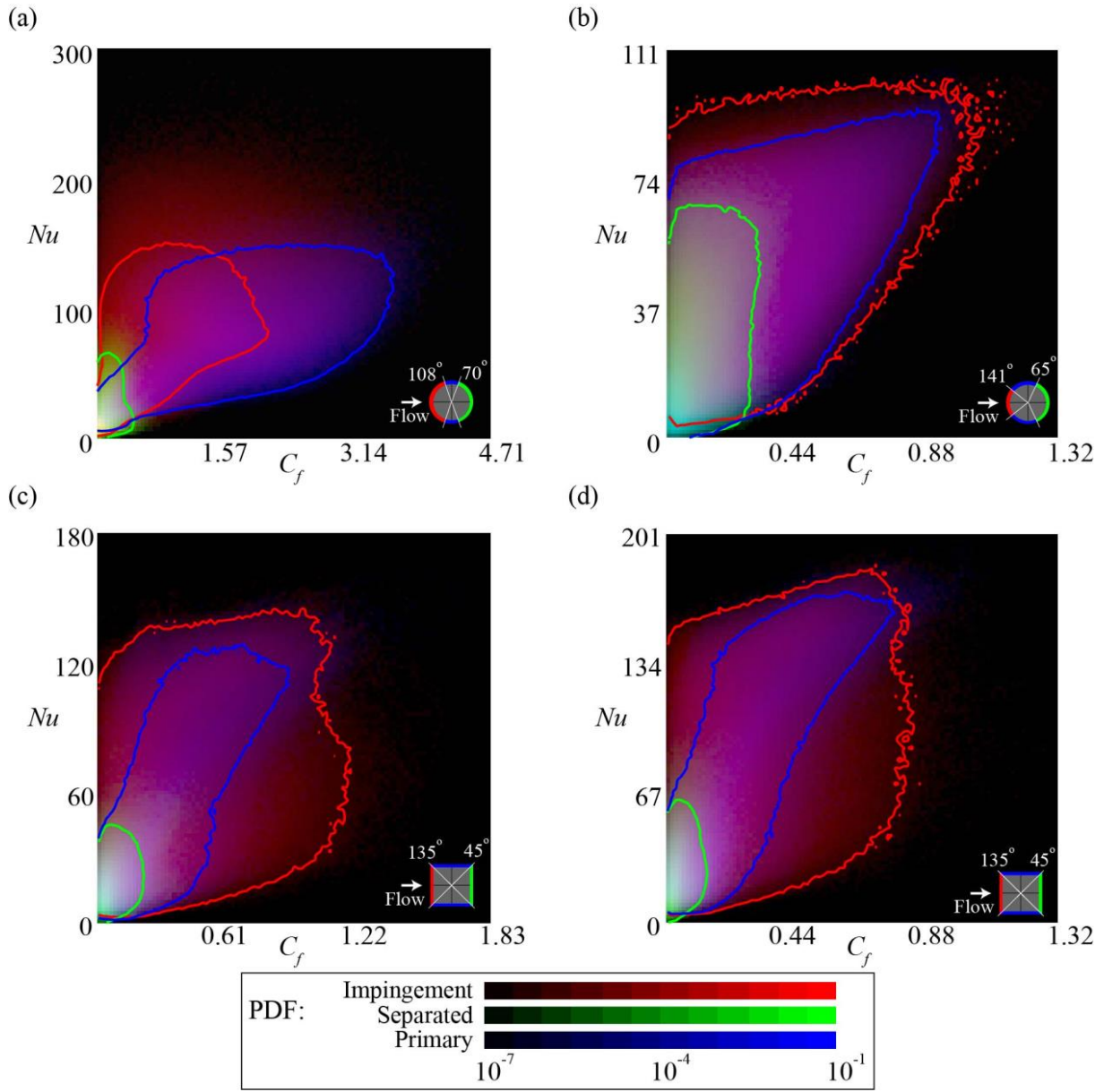

Figure 11: The joint Probability Density Function (PDF) of surface Nusselt number ($Nu$) and skin friction coefficient ($C_f$) for cases (a) A1, (b) A2, (c) A3, and (d) A4. The PDF is plotted for one solid obstacle in the REV in the second row and second column. The solid lines are isolines of a constant PDF value of 0.001 times the maximum value corresponding to the impingement (red), separated (green), and primary (blue) regions. The PDF will be identical for all of the solid obstacles in the REV due to the spatial homogeneity in the porous medium geometry. The individual plots for the different regions are shown in the supplementary information in figures S4-S7.

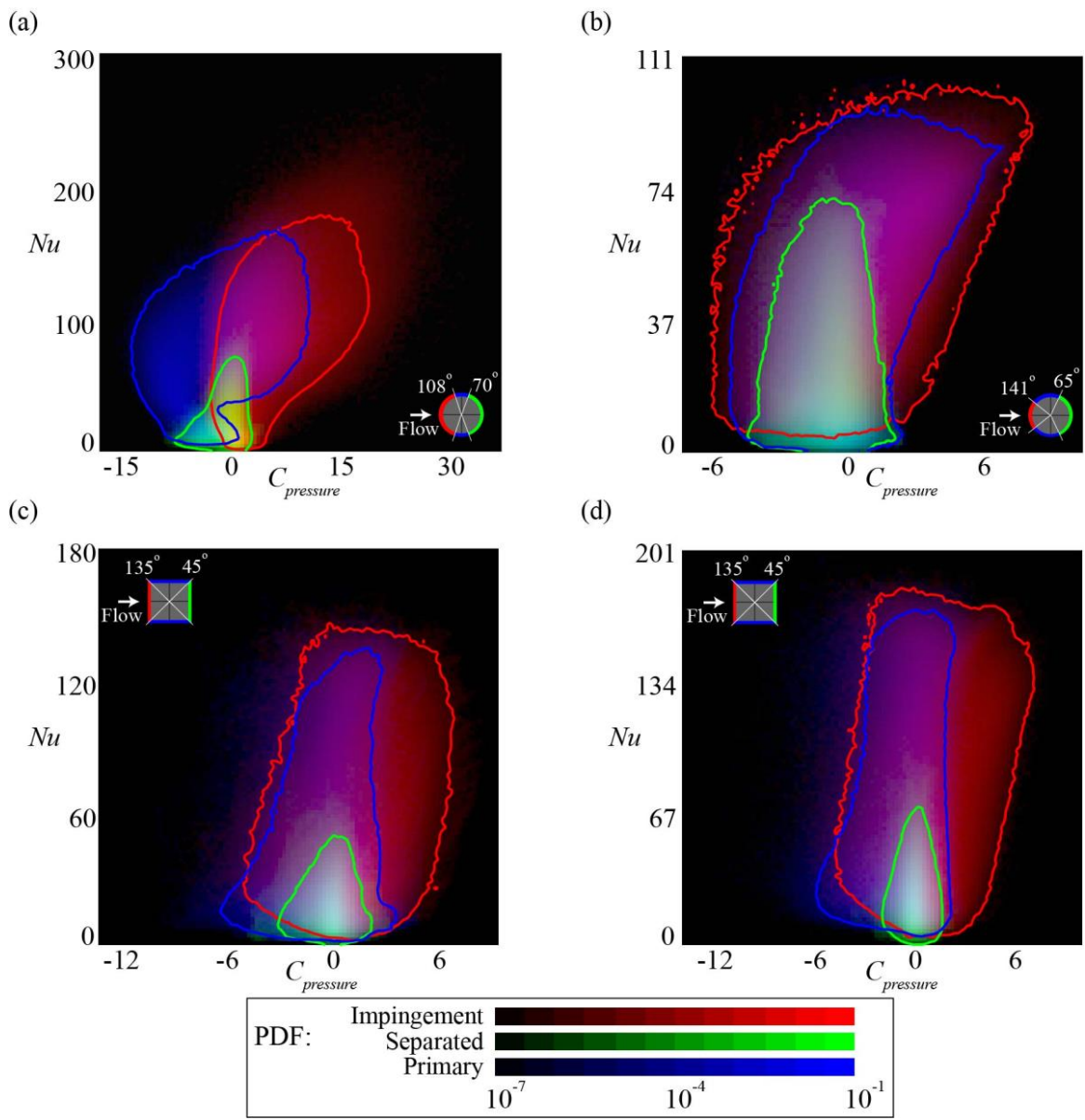

Figure 12: The joint Probability Density Function (PDF) of surface Nusselt number ($Nu$) and coefficient of pressure ($C_{pressure}$) for cases (a) A1, (b) A2, (c) A3, and (d) A4. The PDF is plotted for one solid obstacle in the REV in the second row and second column. The solid lines are isolines of a constant PDF value of 0.001 times the maximum value corresponding to the impingement (red), separated (green), and primary (blue) regions. The individual plots for the different regions are shown in the supplementary information in figures S8-S11.

Consider the PDF distribution on the entire solid obstacle surface. For cases A1, A3, and A4, over 50% of the solid obstacle surface is under low $Nu$, shear, and pressure conditions (table 5). For case A2, a substantial portion (>35%) is under low $Nu$, shear, and pressure conditions. Here, the value of $Nu$ is said to be low if its magnitude is less than the time and surface averaged value. Shear and pressure are said to be low if the magnitude is less than 10% of the maximum value on the solid obstacle surface. This is the reason for the concentration of the PDF distribution at the low $Nu$, shear, and pressure conditions. The contribution of the high PDF areas in figure 11 and figure 12 towards the total heat transfer rate varies considerably across the cases in table 6. In case A1, the bands of high Nu (figure 10(a)) near the stagnation point promote heat transfer even though a majority of the remaining surface area is under the low *Nu* condition. In case A2, the PDF is more distributed due to a lack of a clear distinction between the flow regions. The range of Nusselt numbers in this case is smaller than in the other cases. These factors do not influence the shear as much since the shear layers are formed in case A2 similar to case A1. However, the pressure distribution is affected by the porosity increase from case A1 to A2. Therefore, the high probability density region covers a large portion of figure 12(b). Cases A3 and A4 both have ~50% of the total heat transfer from the high probability regions. This is because the peak *Nu* regions are observed only at the vertices of the square solid obstacle. The remainder of the surface area can be seen to have a consistently low Nu in figures 10(c) and (d). An interesting observation in table 6 is the similarity in the percentage contribution towards heat transfer from low shear and low pressure regions between identical solid obstacle shapes. It is not apparent at this stage whether this result is coincidental and requires further study.

| Case | % of solid obstacle surface area under low *Nu* and $C_f$ conditions | % of solid obstacle surface area under low *Nu* and $C_{pressure}$ conditions |
|------|------|------|
| A1 | 62% | 59% |
| A2 | 45% | 35% |
| A3 | 62% | 52% |
| A4 | 62% | 60% |

Table 5: Percentage of the solid obstacle surface area under low *Nu*, low $C_f$, and low $C_{pressure}$ conditions where $< \overline{Nu_m}$, $C_f/C_{f,max} < 0.1$, $|C_{pressure}|/|C_{pressure}|_{max} < 0.1$.

| Case | % of total heat transfer rate on the portion of the solid obstacle surface with: | | | |
|------|------|------|------|------|
| | $C_f$-*Nu* PDF > 0.1 ($C_f$-*Nu* PDF)$_{max}$ | $C_{pressure}$-*Nu* PDF > 0.1 ($C_{pressure}$-*Nu* PDF)$_{max}$ | $C_f/C_{f,max} < 0.1$ | $|C_{pressure}|/|C_{pressure}|_{max} < 0.1$ |
| A1 | 16% | 19% | 36% | 48% |
| A2 | 22% | 52% | 34% | 49% |
| A3 | 48% | 46% | 59% | 58% |
| A4 | 45% | 42% | 56% | 67% |

Table 6: Percentage of the total heat transfer rate from the solid obstacle surface with high probability density, low $C_f$, and low $C_{pressure}$.

## 4. Conclusions

Microscale vortices play an essential role in turbulent convection heat transfer in porous media. The micro-vortices have a higher core temperature than the primary flow due to flow recirculation inside them. The rotation of the vortices entrains the fluid from the primary flow and raises its temperature by absorbing heat from the solid obstacle wall. This process contributes favorably to the heat transfer. In addition to the primary vortices, there is a pair of secondary vortices that are formed in the streamwise void space at low porosities such as $\varphi =$ 0.50. The secondary vortices are recirculating vortices that are attached to the solid obstacle surface, and they have an adverse effect on heat transfer. This suggests that the dynamic properties of the primary vortices are desirable to increase the heat transfer rate.

The dynamics of flow inside porous media are characterized by two main flow instabilities: the vortex shedding instability and a secondary flow instability. The instabilities are sensitive to the porosity and the solid obstacle shape of the porous medium. Consider a porous medium composed of circular cylinders. At the low porosity of $\varphi = 0.50$, the vortex shedding process is constrained by the streamwise void space. This causes the primary vortices to be localized in the streamwise void space. In comparison, the vortices are not constrained by the streamwise void space at a higher porosity of $\varphi = 0.87$, allowing them to be advected into the primary flow region. As a result, there is a higher concentration of high temperature fluid in the secondary flow region for low porosity when compared to high porosity. This limits the high Nusselt number regions at the low porosity to the vicinity of the stagnation line. At the high porosity, the solid obstacle shape does not influence the mean Nusselt number significantly. This suggests that the solid obstacle shape will not affect heat transfer rate to a significant degree at high porosities. However, the dynamics of heat transfer is influenced by the solid obstacle shape. For square cylinder solid obstacles, the flow separation lines are prescribed to be located near the vertices of the square geometry. As a result, the amplitude of oscillation of the coefficient of drag due to vortex shedding is less for the square solid obstacles when compared to the circular solid obstacles. The amplitude of oscillation of the coefficient of drag due to the secondary flow instability is similar for both the square and circular solid obstacles.

At the macroscale, the FFT spectra of the time series of $Nu_m$, $C_D$, and $C_L$ share similar peaks corresponding to the flow instabilities in each case. The dynamics of heat transfer closely follows the dynamics of the vortex systems. The peaks in the FFT spectra of the $C_D$, $C_L$, and $Nu_m$ correspond to the micro-vortex shedding process and a secondary flow instability, as well as turbulent fluctuations in the microscale flow. The inferences from the flow visualization support these observations as well. This confirms our hypothesis that the time dependent dynamics of the microscale heat convection is characterized by the microscale turbulent structures dynamics.

The magnitude of the macroscale Nusselt number is derived from a highly inhomogeneous distribution of the surface Nusselt number. The inhomogeneity is brought by micro-vortex formation and interaction with the solid obstacles. From the PDF analysis, it is observed that the flow features inside a porous medium: impingement, separated, and primary, introduce unique heat transfer characteristics. The 'impingement' and the 'primary' regions contribute a high Nusselt number due to their proximity to the stagnation point. Meanwhile, the 'separated' region typically has a low Nusselt number as long as the flow has distinct primary and secondary flow regions. A considerable portion (35-60%) of the heat transfer arises from low shear and low pressure conditions on the solid obstacle surface. This is due to the large probability density in these conditions. High Nusselt number areas are always in the vicinity of flow stagnation and cover only a small portion of the solid obstacle surface. We also observe that there are no regions that have low Nusselt number and high shear. The areas having high Nusselt number and high shear exhibit highly unsteady characteristics, which include the change in the location and strength, due to the interaction with the micro-vortices. All of these observations reaffirm the role of the properties of microvortices in heat transfer in porous media.

This includes not only the length scale of the vortex, but more importantly the time scale and the shedding mechanism.

**Acknowledgment**

AVK acknowledges with gratitude the support of the National Science Foundation (award CBET-2042834), the Alexander von Humboldt Foundation through the Humboldt Research Award, and the Extreme Science and Engineering Discovery Environment (XSEDE), which is supported by National Science Foundation grant number ACI-1548562.

**Declaration of interests.** The authors report no conflict of interest.

**Appendix A. Analysis of the REV size convergence**

In this appendix, the convergence of the macroscale flow solution at the chosen REV size is demonstrated. A Reynolds number of 1,000 and porosity of 0.80 are used. The solid obstacles used are circular cylinders (similar to case A2). The REV size is increased from 1*s* to 5*s* in increments of 1*s*. A grid resolution of 0.02*s* is used to perform LES for each of the REV sizes.

The analysis is performed using the following macroscale quantities: the mean applied pressure gradient $\overline{g_1}$, and the macroscale Turbulence Kinetic Energy (TKE) $\langle k_{TKE} \rangle^i$ (figure 12), where the over bar indicates time averaging. The applied pressure gradient determines the drag force on the surface of the solid obstacles inside the porous medium. The macroscale TKE is an indicator of the convergence of the flow dynamics. Both $\overline{g_1}$ and $\langle k_{TKE} \rangle^i$ converge at an REV size of 4*s*. When the REV size is increased from 4*s* to 5*s*, $\overline{g_1}$ and $\langle k_{TKE} \rangle^i$ change only 0.4% and 0.25%, respectively. There is a staggered trend observed in figure 13 depending on whether the REV consists of an odd or even number of solid obstacles. The cause of the staggering is a decoupling between the odd and even number REVs is brought about by the influence of the periodic boundary condition that is imposed. The number of modes of the microscale flow instability that can be present in the domain also plays a role.

The distinction between the odd and even number REVs is virtually non-existent at the REV size of 4*s*. This offers further confirmation that the REV size of 4*s* is adequate for the simulations presented in this paper. The REV size in the *z*- direction is halved, from 4*s* to 2*s*. We found that the turbulence two-point correlation function de-correlates in the *z*- direction in a span of 1*s*. We also observed that the size of the turbulent structures is smaller in the *z*- direction when compared to the *x*- and *y*- directions.

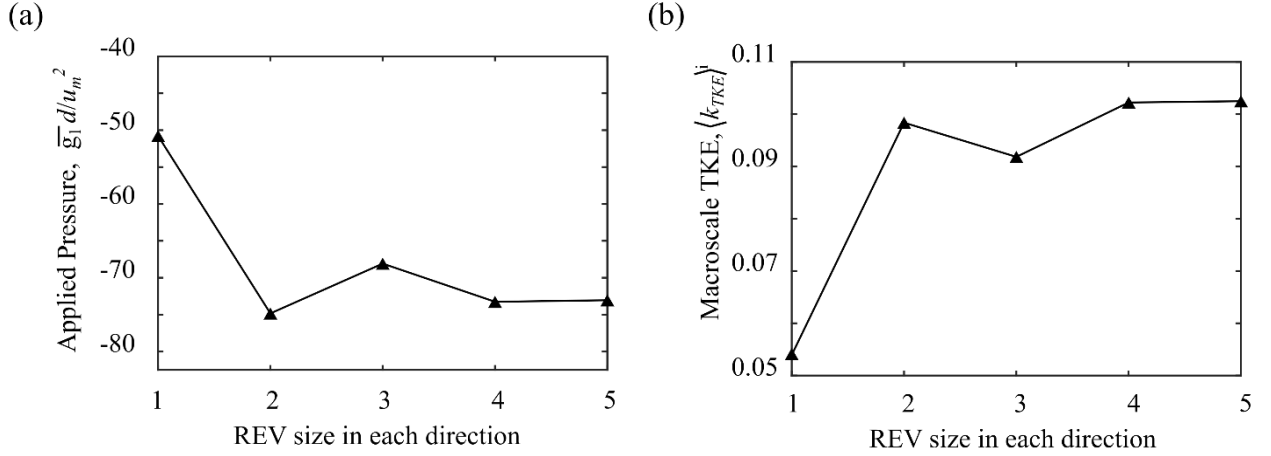

Figure 13: (a) The applied pressure gradient and (b) the macroscale TKE versus REV size at Reynolds number of 1,000 and porosity of 0.80.

**Appendix B. Sensitivity study of the Nusselt number distribution to the characteristic temperature difference**

In this appendix, we demonstrate that the Nusselt number distribution is independent of the characteristic temperature difference ($\Delta T$) near the chosen value of $\Delta T = 30$ K. The representative case used in this study consists of circular cylinder solid obstacles forming a porous medium with a porosity of 0.87. The representative elementary volume (REV) consists of 2 solid obstacles each in the *x*- and *y*- directions. Turbulent flow is simulated using LES at a Reynolds number of 300 for three values of $\Delta T$ – 15 K, 30 K and 60 K. The distributions of the time-averaged Nusselt number on the surface of the first solid obstacle in the REV (row 1, column 1) are shown in figure 13. The Nusselt number distributions are identical for all three values of $\Delta T$ simulated here. The magnitude of the heat flux on the solid obstacle surface is different in each case, but it scales linearly with $\Delta T$ resulting in identical heat transfer coefficients for the three cases. Following this result, we are using the characteristic temperature difference $\Delta T = 30$ K for the remainder of the simulations in this paper.

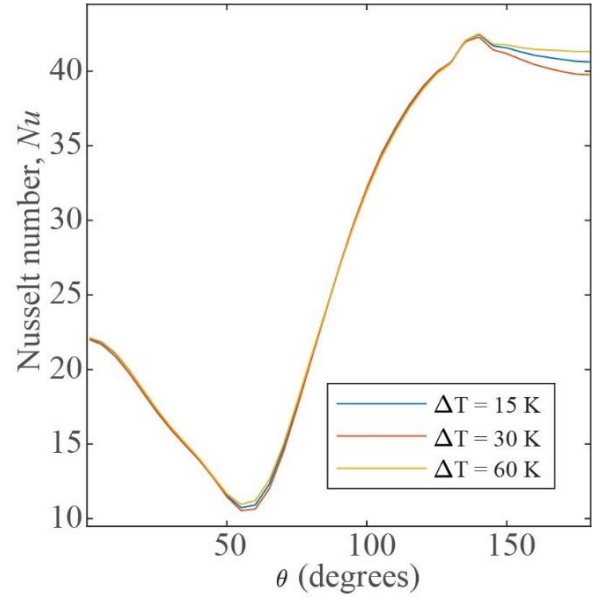

Figure 14: The distributions of the time-averaged Nusselt number on the surface of the solid obstacle at row 1, column 1 in the REV for three values of the characteristic temperature difference $\Delta T$ – 15 K (red), 30 K (green) and 60 K (blue).

**Appendix C. Phase difference in surface forces between different solid obstacles**

In this appendix, we use the coefficient of lift ($C_L$) to demonstrate the phase difference in the surface forces between different solid obstacles. The same phenomenon can be observed for other solid obstacle surface forces such as the drag.

For $Re_p = 300$, the von Karman instability is observed for the flow around each solid obstacle. The coefficient of lift on each solid obstacle is in phase for all the solid obstacles in the same column. Figure 14 illustrates $C_L$ over time for the case of $Re_p = 300$, $\varphi = 0.87$ with circular cylinder solid obstacles (case A2) of the same column. We can observe in movie 2 that any sway in $C_L$ out of phase will be corrected by the flow around neighboring solid obstacles of the same column. The coefficients of lift of the solid obstacles in the same row exhibit phase difference. Figure 14 shows that $C_L$ peaks at different times for the solid obstacles in the same row. When $C_L$ for each solid obstacle in the REV is summed, the amplitude of the resultant coefficient of lift is less than the sum of the individual amplitudes. There exists an interference between the individual contributions to $C_L$ of the solid obstacles due to the phase difference. Therefore, we do not use the resultant forces over the REV to perform our analysis.

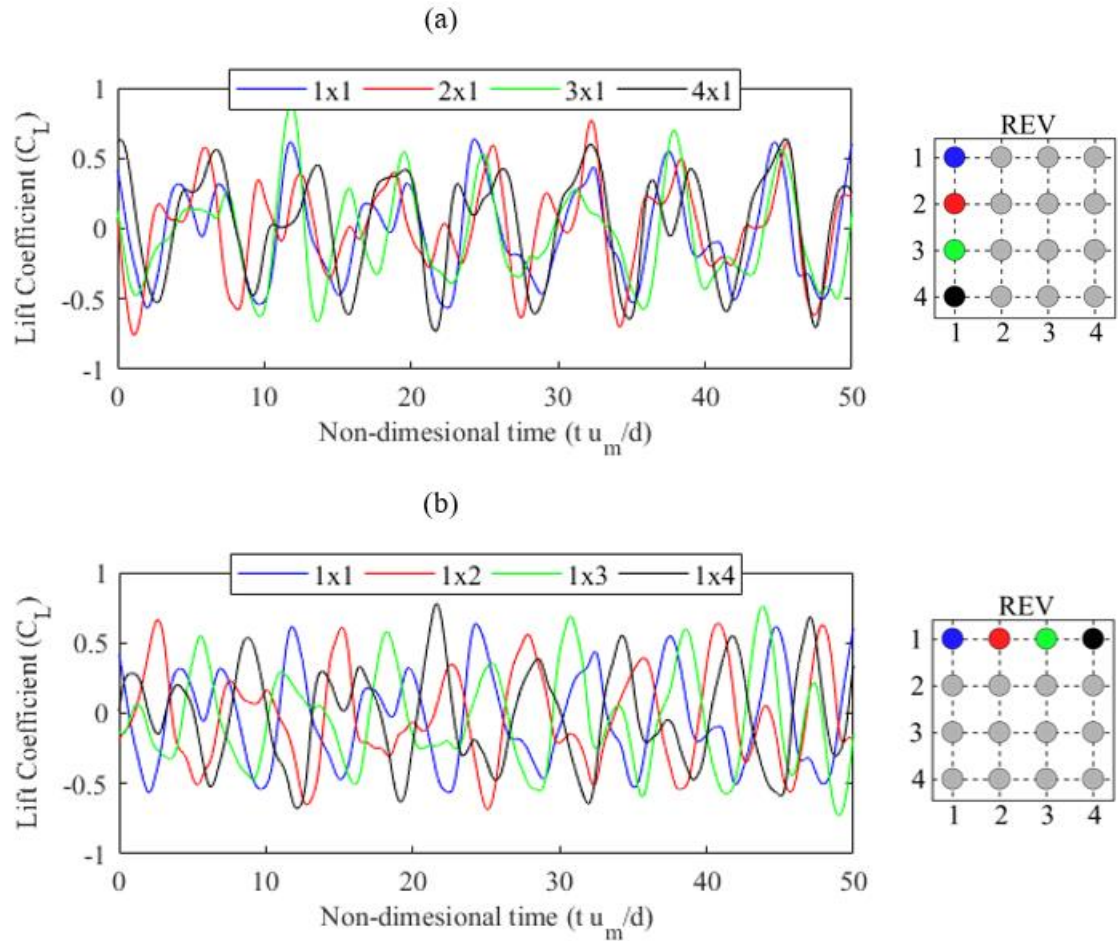

Figure 15: Change of lift coefficient ($C_L$) over time for solid obstacles in (a) the 1st column and (b) the 1st row for case A2 ($Re_p = 300$, $\varphi = 0.87$ with circular cylinder solid obstacles).